\documentclass[10pt]{article}
\usepackage[margin=1in]{geometry}                % See geometry.pdf to learn the layout options. There are lots.
\usepackage[parfill]{parskip}    % Activate to begin paragraphs with an empty line rather than an indent
\usepackage{graphicx}
\usepackage{amssymb}
\usepackage{epstopdf}
\usepackage{amsmath}
\usepackage{hyperref}
\usepackage{mathrsfs}
\usepackage{paralist}
\usepackage{varioref}
\usepackage{ulem}
\usepackage[compress]{cite}

\hypersetup{
    bookmarks=true,         % show bookmarks bar?
   pdfstartview={FitH},    % fits the width of the page to the window
   colorlinks=true,       % false: boxed links; true: colored links
   linkcolor=black,          % color of internal links
   citecolor=black,        % color of links to bibliography
   filecolor=black,      % color of file links
   urlcolor=black           % color of external links
}

\DeclareFontFamily{OT1}{pzc}{}
\DeclareFontShape{OT1}{pzc}{m}{it}%
             {<-> s * [0.900] pzcmi7t}{}
\DeclareMathAlphabet{\mathscr}{OT1}{pzc}%
                                 {m}{it}
\labelformat{section}{Section #1} 
\labelformat{subsection}{Section #1} 
\labelformat{subsubsection}{Section #1}
\labelformat{equation}{Eq.~(#1)} 
\labelformat{figure}{Fig.~#1} 
\labelformat{subfigure}{Fig.~\thefigure#1} 
\labelformat{table}{Tab.~#1}
\labelformat{appendix}{Appendix #1}
\newcommand{\be}{\begin{equation}}
\newcommand{\ee}{\end{equation}}
\newcommand{\bea}{\begin{eqnarray}}
\newcommand{\eea}{\end{eqnarray}}

\newcommand{\nn}{\nonumber}

\newcommand{\sqg}{\sqrt{-g}}
\newcommand{\sqh}{\sqrt{h}}
\newcommand{\mLq}{\mathcal{L_{\textrm{quad}}}}
\newcommand{\mLs}{\mathcal{L_{\textrm{sur}}}}
\newcommand{\mLm}{\mathcal{L_{\textrm{m}}}}
\newcommand{\mH}{\mathcal{H}}
\newcommand{\Lq}{L_{\textrm{quad}}}
\newcommand{\Ls}{L_{\textrm{sur}}}

\newcommand{\df}{\delta}
\newcommand{\vbar}{\left\vert\vphantom{\int}\right.}

\newcommand{\al}{\alpha}
\newcommand{\bt}{\beta}

\def\LL{Lanczos-Lovelock}
\title{Structure of the Gravitational Action and its relation with Horizon Thermodynamics
and Emergent Gravity Paradigm}
\author{ 
 {\bf {\normalsize Krishnamohan Parattu,}$
$\thanks{E-mail: krishna@iucaa.ernet.in}} \, 
{\bf {\normalsize Bibhas Ranjan Majhi}$
$\thanks{E-mail: bibhas@iucaa.ernet.in}} \, and
{\bf {\normalsize T. Padmanabhan}$
$\thanks{E-mail: paddy@iucaa.ernet.in}}\\
 {\normalsize IUCAA, Post Bag 4, Ganeshkhind,}
\\{\normalsize Pune University Campus, Pune 411 007, India}
\\[0.3cm]
}

\date{\today}                                           % Activate to display a given date or no date

\begin{document}
\maketitle

%%%%%%%%%%%%%%%%%%%%%%%%%%%%%%%%%%%%%%%%%%%%%%%%
\begin{abstract}
If gravity is an emergent phenomenon, as suggested by several recent results, then the structure of the action principle for gravity should encode this fact. With this motivation we study several features of the Einstein-Hilbert action and establish direct connections with horizon thermodynamics. We begin by introducing the concept of holographically conjugate variables (HCVs) in terms of which the surface term in the action has a specific relationship with the bulk term. In addition to $g_{ab}$ and its conjugate momentum $\sqg M^{cab}$, this procedure allows us to (re)discover  and motivate strongly the use of $f^{ab}=\sqrt{-g}g^{ab}$ and its conjugate momentum $N^c_{ab}$. The gravitational action can then be interpreted as a momentum space action for these variables. We also show that many expressions in classical gravity simplify considerably in this approach. For example, the field equations can be written in the form  $\partial_cf^{ab}=\partial \mathcal{H}_g/\partial N^c_{ab},\partial_cN^c_{ab}=-\partial \mathcal{H}_g/\partial f^{ab}$ (analogous to Hamilton's equations) for a suitable Hamiltonian $\mathcal{H}_g$, if we use these variables. More importantly,  the variation of the surface term, evaluated on any null surface which acts a local Rindler horizon can be given a direct thermodynamic interpretation. The term involving the variation of the dynamical variable leads to $T\delta S$ while the  term involving the variation of the conjugate momentum leads to $S\delta T$. We have found this correspondence only for the choice of variables $(g_{ab}, \sqg M^{cab})$ or $(f^{ab}, N^c_{ab})$. We use this result to provide a direct thermodynamical interpretation of the boundary condition in the action principle, when it is formulated in a spacetime region bounded by the null surfaces. We analyse these features from several different perspectives and provide a detailed description, which offers insights about the nature of classical gravity and emergent paradigm.
\end{abstract}

%#######################################################################
\section{Introduction}
The curious connection between gravity and thermodynamics first came to light with the work of Bekenstein \cite{Bekenstein:1972tm,Bekenstein:1973ur,Bekenstein:1974ax}, which ascribed to a black hole an entropy proportional to the surface area of its horizon. Soon, it was discovered that black hole horizons possess temperature as well \cite{Hawking:1974sw,Hawking:1976de}. In the four decades since then, the intriguing connection between gravity and thermodynamics has been steadily becoming stronger (see e.g., \cite{Wald:1999vt,Padmanabhan:2003gd} ).

One paradigm which has emerged from this connection considers the dynamics of gravity as not of fundamental nature, but emergent from the dynamics of a more fundamental theory, just as the thermodynamics of a material system emerges out of the basic dynamics of its molecules. (For a recent review, see \cite{Padmanabhan:2009vy,Padmanabhan:2012qz}; for other work similar in spirit, see e.g.,\cite{Sakarov,Hu:1996gk,Jacobson:1995ab,Barcelo:2005fc,Volovik}). 
This paradigm has obtained support from the following results:
\begin{enumerate}[a)]                                                           \item Gravitational field equations in a wide class of theories -- more general than Einstein gravity -- lend themselves to a thermodynamical interpretation \cite{Padmanabhan:2002sha, Kothawala:2009kc, Padmanabhan:2009jb};
\item Gravitational equations of motion can be obtained from thermodynamical extremum principles  \cite{Padmanabhan:2007en, Padmanabhan:2007xy};
\item It has been possible to obtain the density of microscopic degrees of freedom through equipartition arguments \cite{Padmanabhan:2009kr,Padmanabhan:2010xh};
\item The action functional for gravitation in a class of theories has a thermodynamic interpretation \cite{Padmanabhan:2002sha, Padmanabhan:2004fq, Mukhopadhyay:2006vu, Kolekar:2010dm};
\item Einstein's equations reduce to Navier-Stokes equations of fluid dynamics in any spacetime when projected on a null surface \cite{Padmanabhan:2010rp, Kolekar:2011gw}. This generalizes previous results for black hole spacetime \cite{Damour:1979, Damour:1982, Thorne}.
\item The euclidean path integral of the gravitational action interpreted as a partition function has provided expressions for free energy, energy and entropy  \cite{Kolekar:2011bb} in the \LL\ models generalizing  previously known results \cite{Padmanabhan:2002sha,Gibbons:1976ue}.
\end{enumerate}
These results show that gravity could indeed be an emergent phenomenon, similar to elasticity or fluid mechanics.

Conventionally, however, one treats gravity like any other field and its dynamics is obtained from a standard action principle. The fact that the theory obtained by such a variational principle possesses an emergent description strongly suggests that {\it the action functional itself must encode this information}. The main purpose of this paper is to explore several aspects of gravitational action principle from the emergent gravity perspective and unravel these features to the extent possible.

In fact, it is well-known that general relativity has some peculiarities which introduces special difficulties (not encountered in other field theories, e.g. non-Abelian gauge theories)  when we try to derive the Einstein equations from an action principle (see e.g., Chapter 6 of \cite{gravitation}). There does exist a scalar action, the Einstein-Hilbert action, which, when added to the matter action, can be varied with respect to the metric to obtain the Einstein equations. Based on the usual theoretical prejudice \cite{Ostrogradsky:1850}, the equations of motion which are second order in derivatives of the coordinates, are obtained from an action which is quadratic in the first derivatives of the dynamical variables. The generally covariant Lagrangian for gravity, however, is forced to be at least second order in derivatives of the metric since any scalar made of the metric and its first derivatives will have some constant value in the local inertial frame and, by virtue of being a scalar, has the same constant value in any other frame. The usual choice, $R$, has a special structure (viz. linearity in second derivatives) which allows us to obtain the equations of motion which are only second order in the derivatives of the metric, if (and only if) we fix metric \textit{and} its derivatives on the boundary. This is possible  because we can separate out the the second derivatives in the Lagrangian into a total derivative which becomes a surface term in the action. Its variation does not contribute to the equations of motion if we set the variations of the metric \textit{and} its derivatives to be zero on the boundary.

The main conceptual difficulty with this program --- which makes gravitational action different from those in other field theories --- is that the  we need to fix both the dynamical variable and its derivative at the boundary to obtain the equations of motion. There are some issues with this procedure. Suppose that the spacetime region  we are considering is the region between two spacelike surfaces. Assume that all quantities go to zero at spatial infinites. If we now fix the metric and its derivative on the earlier time-slice, the Einstein equations would give us the corresponding values on the latter time-slice. Thus, we do not really have the freedom to fix arbitrary values for the metric and its derivatives on the boundaries. Also, setting our eyes on a quantum theory, we would not want to fix both the dynamical variable and its derivative at the same spacelike hypersurface.
Another option would be to add a term to the action, the variation of which will cancel the terms with variation of the first derivative. A well-known example is the Gibbons-Hawking-York counterterm \cite{York:1972sj,Gibbons:1976ue} though it is not unique \cite{Charap:1982kn}. But this entire approach appears a little contrived and the resulting action, for the Gibbons-Hawking-York case, although generally covariant becomes foliation dependent. 

There is, however, another aspect to this issue. If we use $g_{ab}$ as the dynamical variables (with an associated  canonical momenta), then it turns out that the surface variation \textit{involves only the variation of the canonical momenta}. This is a nontrivial result and arises \textit{only because} the surface and bulk terms of the Hilbert Lagrangian are connected in a peculiar manner \cite{Padmanabhan:2004fq}. This, in turn, implies that one can treat the Hilbert action as a momentum space action and the equations of motion can be obtained by fixing the canonical momenta at the boundary. Thus, gravity may be better dealt with in the space of the canonical momenta corresponding to the metric rather than in the space of the metric.
Further, we will demonstrate that this program will work with the components of $g_{ab}$ as variables, but will fail with $g^{ab}$ as variables. Although $g^{ab}$ is a failure, we shall find that the components of the corresponding tensor density, $f^{ab}=\sqg g^{ab}$, would also serve our purpose. 

The variables $f^{ab}=\sqg g^{ab}$ have been used in classical literature \cite{Eddington:1924,Schrod:1950,Einstein:1955ez,Einstein-Kauffman} but somehow does not seem to have attracted sufficient attention in recent years. One of the purposes of this paper is to advertise nice features of these variables and some pedagogical results.
After a slight digression on how various expressions simplify on being expressed in terms of $f^{ab}$ and the corresponding canonical momenta, we look more closely at the result that an integral of the surface term in Einstein-Hilbert action over a horizon gives the entropy \cite{Padmanabhan:2002sha, Padmanabhan:2004fq, Mukhopadhyay:2006vu, Kolekar:2010dm,Padmanabhan:2012bs,Majhi:2013jpk} of the horizon. The results have been obtained for static metrics in which the action will be proportional to the  range of time integration $\tau$ due to the static nature of the metric. The integrals of the surface Lagrangian over the horizon will give $\tau TS$, where $\tau$ is the range of the time integration, $T$ is the temperature and $S$ is the entropy. Usually we work in Euclidean sector where the natural choice for $\tau$ is  the inverse temperature $\beta$, giving $\tau TS=S$. Alternatively, one can define the surface Hamiltonian \cite{Padmanabhan:2012qz,Padmanabhan:2012bs,Majhi:2013jpk} by $H_{\textrm{sur}} = -(\partial A_{\textrm{sur}}/\partial\tau) = TS$ and concentrate on just the integration over the co-ordinates transverse to the horizon. A variation  of this integral would provide us with $T \df S +  S \df T$.  We will show that the variation of the surface term in the Einstein-Hilbert action for static metrics allows us to identify the $T \df S $ term as arising from  the variation of the dynamical variable $g_{ab}$ or $f^{ab}$ and the  $S \df T$ term arises from  the variation of the corresponding canonical momenta in both cases. (In contrast, such a nice separation does \textit{not} work if, e.g., we use $g^{ab}$ as dynamical variables). 

This result extends to a very general class of null surfaces, which act as horizons for local Rindler observers and suggests a simple thermodynamical interpretation of the variational principle. In any region bounded by null surfaces, one can introduce local Rindler observers who perceive patches of the null surfaces as horizons and attribute  temperatures to them. Using this, we can  reformulate the boundary condition for the gravitational action principle as equivalent to keeping the temperature constant on these null surfaces. We believe this offers some insight into the thermodynamic interpretation of action principle.

The outline of the paper is as follows. In \ref{prelims}, we review some well-known results on the structure of the Einstein-Hilbert action to set the stage. In \ref{good}, we show  why the variables $g_{ab}$ and $f^{ab}=\sqg g^{ab}$ are preferable over $g^{ab}$ in choosing the dynamical variables for formulating an action principle for general relativity. \ref{canonvar} provides some useful relations between known expressions in relativity and our canonical variables. In particular, the field equations can be written in the form  $\partial_cf^{ab}=\partial \mathcal{H}_g/\partial N^c_{ab},\partial_cN^c_{ab}=-\partial \mathcal{H}_g/\partial f^{ab}$ (analogous to Hamilton's equations) for a suitable Hamiltonian $\mathcal{H}_g$, if we use these variables. \ref{therm} discusses the variation of the surface term in the action and explores its relation with the canonical variables. Finally, we present our conclusions in \ref{concl}. 

The conventions used in this paper are as follows. We use the metric signature $(-,+,+,..,+)$. The fundamental constants $G$, $\hbar$ and $c$ have been set to unity. The Latin indices run over all space-time indices while Greek indices will be used for purely spatial indices. The tensor density corresponding to a tensor will be denoted by the corresponding letter in calligraphic font. For example, the Lagrangian density will be denoted by $\mathcal{L}=\sqg L$. 

%#######################################################################
\section{Preliminaries: The Structure of the Einstein-Hilbert Action}\
\label{prelims}
In this section, we shall rapidly review some ideas and relations known in the literature, in order to set the stage. As we said, there is an inherent difficulty in trying to build an action principle for general relativity from which the gravitational equations of motion can be derived, which sets it apart from all other theories. The dynamical equations of motion are not expected to contain derivatives of the variable of order greater than two. This normally requires the action to contain not more than  the first derivatives of the dynamical variables. But in general relativity, we would like the Lagrangian to be a scalar and any such scalar built out of the first derivatives will have some constant value in the local inertial frame, and by virtue of being a scalar will have the same constant value in any other frame. The traditional way of dealing with this situation is to construct an action consisting of the metric and its first and second derivatives and then arrange matters such that the equations of motion are only second order in the derivatives of the metric. In fact, this demand alone is enough to uniquely identify the action in $D=4$ dimensions. This action is the Einstein-Hilbert action, given by
\begin{equation}
16\pi A_{\rm EH}= \int_\mathcal{V} d^4x\, \mathcal{L}_{\rm EH} =\int_\mathcal{V} d^4x\sqrt{-g}\, R.
\label{A_EH}
\end{equation} 
with $L_{EH}=R$ in our notation. It is useful to define a quantity $Q_{a}^{\phantom{a}bcd}$ and write the Lagrangian density in an equivalent form as:
\begin{equation}
\mathcal{L}_{\rm EH}\equiv \sqrt{-g} Q_a^{\phantom{a}bcd}R^a_{\phantom{a}bcd}; \qquad Q_a^{\phantom{a}bcd}=\frac{1}{2}(\delta^c_ag^{bd}-\delta^d_ag^{bc}).
\label{lisrq}
\end{equation}
The advantage of this form is that it is readily generalized to gravitational theories in more than $4$ dimensions where our conditions allow other terms in the action in addition to the Einstein-Hilbert term (see e.g., Chapter 15 in \cite{gravitation}, \cite{Padmanabhan:2013xyr}).  It is a well-known result that $\mathcal{L}_{\rm EH}$ can be decomposed into a bulk term, which is quadratic in the derivatives of the metric, and a surface term which contains all the second derivatives (see \cite{Eddington:1924,Schrod:1950}). The variation of the bulk term alone can furnish the equations of motion (as explicitly shown in e.g., chapter 6 of \cite{gravitation}) while  the surface term, when integrated over a horizon, is related to the entropy of the horizon. The decomposition into bulk and surface terms is given by 
\begin{align}  
\mathcal{L}_{\rm EH} = \mathcal{L}_{\rm quad}+\mathcal{L}_{\rm sur}, 
\label{decomp}
\end{align}
where we have defined the bulk Lagrangian density and the surface Lagrangian density respectively as
\begin{equation}
\mathcal{L}_{\rm quad}\equiv2 \sqrt{-g}Q_a^{\phantom{a}bcd}\Gamma^a_{dk}\Gamma^k_{bc};\qquad
\mathcal{L}_{\rm sur}\equiv2\partial_c\left[\sqrt{-g}Q_a^{\phantom{a}bcd}\Gamma^a_{bd}\right].
\label{bulksur}
\end{equation}
We will also  require the expression  for the bulk Lagrangian in terms of the derivatives of the metric. This  is given by:
\begin{equation}
L_{\rm quad}=\frac{1}{4}M^{abcijk}\partial_a g_{bc} \partial_i g_{jk}, 
\label{derivLq}
\end{equation}
where we have defined a $6-$indexed tensorial object $M^{abcijk}$, given by:
\begin{align}
M^{abcijk}={}&g^{ai}g^{bc}g^{jk}
          -\frac{1}{2}(g^{ai}g^{bj}g^{ck}+g^{ai}g^{cj}g^{bk}) \nonumber \\
          {}&+\frac{1}{2}(g^{ak}g^{bj}g^{ci}+g^{aj}g^{bk}g^{ci})
          +\frac{1}{2}(g^{ak}g^{cj}g^{bi}+g^{aj}g^{ck}g^{bi}) \nonumber \\
          {}&-\frac{1}{2}(g^{ak}g^{bc}g^{ij}+g^{aj}g^{bc}g^{ik}
          +g^{ib}g^{jk}g^{ac}+g^{ic}g^{jk}g^{ab}).
\end{align}
This has the following symmetry properties. It is symmetric in $b$, $c$ and $j$, $k$ and also under exchange of the index triplets ($abc$) and ($ijk$).
\ref{derivLq} allows us to obtain the canonical momenta corresponding to $\mathcal{L}_{quad}$ as 
\begin{equation}
 \sqg M^{abc}= \frac{\partial \sqrt{-g}L_{\rm quad}}{\partial (\partial_a g_{bc})}= \frac{1}{2} \sqg\ M^{abcijk}\partial_i g_{jk}.
 \label{M3}
\end{equation}
Note that $M^{abc}$ is not a tensor. Nevertheless, we can define a rule for raising and lowering of indices as in the case of tensors and lower the last two indices to define:
\begin{equation}
 M^{a}_{bc}\equiv g_{bd}g_{ce}M^{ade} =-\frac{\partial L_{\rm quad}}{\partial (\partial_a g^{bc})}. \label{Mudd}
\end{equation}
Thus, $\sqg M^{a}_{bc}$ is the negative of the canonical momentum corespondent to $g^{bc}$. The negative sign in the last term arises because $\partial_a g^{bc}=-g^{bd}g^{ce}\partial_a g_{de}$. 

As we have stated already, the equations of motion can be derived from the bulk term alone.
Once the equations of motion are obtained, we can find solutions to them, including black hole solutions, say, without ever bringing into discussion the surface term. If we now evaluate the 
surface term of the action, on the horizon, it will reproduce the entropy of the horizon  in the Euclidean sector. (In general it gives $\tau TS$ where $\tau$ is the range of integration; we get $S$ when $\tau=\beta$, the inverse temperature.) 
The fact that the surface term, which is not supposed  to know anything about the equations of motion, gives the entropy when integrated over a horizon demands an explanation. \textit{We stress that, this result --- in fact --- is a direct hint that gravitational action principle contains information about horizon thermodynamics.} 

The algebraic answer to this question lies in realizing that the bulk term $\mathcal{L}_{\rm quad}$ and the surface term $\mathcal{L}_{\rm sur}$ are not independent, but related to each other by the relation
\begin{eqnarray}
\mathcal{L}_{\rm sur}&=&-\left[\partial_c\left(g_{ab}\frac{\partial \mathcal{L}_{\rm quad}}{\partial (\partial_c g_{ab})}\right)\right]. 
\label{holorel}
\end{eqnarray}
Since it relates a quantity on the surface with a quantity in the bulk, this relation has been termed ``holography'' in the past \cite{Padmanabhan:2004fq,Mukhopadhyay:2006vu,Kolekar:2010dm}. It is the same  $-\partial \left( pq\right)$ structure of the surface term that allows us to derive the equations of motion by fixing the canonical momentum at the boundary rather than the metric \cite{Mukhopadhyay:2006vu}. We shall discuss this issue in the next section.

Using the canonical momenta $M^{abc}$ as defined in \ref{M3}, we can rewrite \ref{holorel} as
\begin{eqnarray}
 \mathcal{L}_{\rm sur}=-\left[\partial_c\left(\sqrt{-g}g_{ab}M^{cab}\right)\right]
        \equiv\partial_c \left( \sqg V^{c} \right), \label{surVc}
\end{eqnarray}
where we have defined a one-indexed non-tensorial object $V^{c}$ given by
\begin{equation}
 V^{c}\equiv -g_{ab} M^{cab}=g^{ik} \Gamma^{c}_{ik}-g^{ck}\Gamma^{m}_{km}=2 Q_{a}^{\phantom{a}bcd}\Gamma^{a}_{bd}=2 Q^{ijkc}\partial_i g_{jk}=-\frac{1}{g}\partial_b(g g^{bc}).
 \label{Vc}
\end{equation}
Using the $V^{c}$ thus defined, we can write down the following expression for the canonical momentum $M^{abc}$:
\begin{equation}
M^{abc}=g^{bd}g^{ce} \Gamma^{a}_{de} - \frac{1}{2}g^{bd} g^{ac} \Gamma^{e}_{de} - \frac{1}{2} g^{cd} g^{ab} \Gamma^{e}_{de}  - \frac{1}{2}g^{bc}V^{a}. \label{M_in_GammaV}
\end{equation}
Having thus set up the initial framework, we shall now turn to the discussion of how certain ``holographically conjugate'' variables (HCVs) are more suited for the variation of the gravitational action.
%#######################################################################
\section{Holographically Conjugate Variables}
\label{good}
\subsection{Variation of the Action and the Holographically Conjugate Variables}

Let us now consider the variation of the Einstein-Hilbert action. The variation of the  Lagrangian in \ref{decomp} is (see Chapter 6 in \cite{gravitation}) given by
\begin{eqnarray}
\df \mathcal{L}_{\textrm{EH}}= \df (\{\mathcal{L}_{\rm quad}\}+\{\mathcal{L}_{\rm sur}\})
&=& \{\sqg G_{ab} \df g^{ab} - \partial_c[\sqg M^{c}_{ab} \df g^{ab}]\} -\df\{ \partial_c(\sqg g^{ab} M^{c}_{ab})\} \nonumber \\
&=& \sqg G_{ab} \df g^{ab} -2 \partial_c[\sqg M^{c}_{ab} \df g^{ab}] - \partial_c[g^{ab} \df( \sqg M^{c}_{ab}) ] \label{varRMgu}
\end{eqnarray}
where $M^{c}_{ab}$ is as defined in \ref{Mudd}. 
Note that the surface variation has two parts, one part arising out of the variation of the metric while the other part contains the variation of  $\sqg M^{c}_{ab}$, which is the negative of the canonical momentum corresponding to $g^{ab}$. Hence, in order to obtain the equations of motion we have to fix both $g^{ab}$ and its corresponding canonical momentum at the boundaries. But as we have argued in the introduction, there are some fundamental difficulties with this program.

On the other hand, if we write the variation in terms of $g_{ab}$, then a crucial sign flip leads to the cancellation of two terms (which added together in the previous case) and we find
\begin{equation}
\delta(\sqrt{-g}R)=-\sqrt{-g}G^{ab}\delta g_{ab}-\partial_{c}[g_{ik}\delta(\sqrt{-g} M^{cik})],  \label{varRMg1}
\end{equation}
Thus we only need to fix the canonical momenta $\sqrt{-g} M^{cik}$, corresponding to $g_{ik}$,  at the boundary to obtain the equations of motion. So, some of the concerns raised in the introduction about the variational principle in general relativity can be addressed by treating the action principle as a momentum space action and using the covariant components of the metric, $g_{ab}$ as dynamical variables. [The addition of the term $-\partial(pq)$ has a direct interpretation in quantum theory \cite{Padmanabhan:2004fq}. Usually, the propagator $G(q_2,q_1)$ for the dynamical variable $q$ can be obtained from a path integral using the action built from a quadratic Lagrangian $L_q(q,\partial q)$. The propagator $G(p_2,p_1)$  in momentum representation can be similarly obtained from a path integral using the action built from $L_q(q,\partial q)-\partial(pq)$.]

Since we know that $M^{cik}$ is made of the metric and its first derivatives, it is natural to ask how the surface variation term in \ref{varRMg1} looks like in terms of variations of the metric and the affine connection, in the spirit of Palatini. (We shall discuss the Palatini variational principle in \ref{sec:Palatini}.) The required expression is
\begin{equation}
-\partial_{c}[g_{ik}\delta(\sqrt{-g} M^{cik})]=- \partial_{c}[2\sqrt{-g}g^{bk}\delta (Q^{cd}_{be}\Gamma^{e}_{dk})]\equiv - \partial_{c}[\sqrt{-g}g^{bk}\delta N^{c}_{bk}] \label{vargMvarfN}~,
\end{equation}
where $Q^{cd}_{ab}=\frac{1}{2}(\df_{a}^{c}\df_{b}^{d}-\df_{a}^{d}\df_{b}^{c})$ is obtained by lowering an index in $Q_{a}^{bcd}$ as defined in \ref{lisrq}.  We have introduced here a new object, 
\begin{eqnarray}
N^{a}_{bc} = -\Gamma^{a}_{bc}+\frac{1}{2}(\Gamma^d_{bd}\delta^{a}_{c}+\Gamma^d_{cd}\delta^{a}_{b})=  Q^{ad}_{be}\Gamma^{e}_{cd}+ Q^{ad}_{ce}\Gamma^{e}_{bd}~, \label{NGamma}
\end{eqnarray}
which is made purely from the affine connection and does not involve the metric. The reason for introducing this object is not just aesthetics as will be clear later, in \ref{subsec:fN}, along with the reason for the demand for symmetrisation in the lower indices. 

Let us compare \ref{vargMvarfN} with \ref{varRMg1} and \ref{varRMgu}. We obtained some level of simplification in going from \ref{varRMgu} to \ref{varRMg1}. But $\sqg M^{cab}$, in spite of its simple interpretation as the canonical momentum, is a complicated object in terms of the metric and its derivatives (\ref{M3}) or in terms of the metric and the affine connection (\ref{M_in_GammaV}).  Hence, it is a pleasant surprise to find that the surface variation turns out to be a pure variation of the affine connection. As a consequence, we can let the metric be arbitrary at the boundary and fix just the connection at the boundary to obtain the equations of motion. If we consider the metric and the affine connection as two different fields \`{a} la Palatini, then this is a perfectly acceptable situation, although a little peculiar since we need to fix only one field out of the two that we are considering. The reason for this can be traced to the fact that the Einstein-Hilbert Lagrangian can be expressed in terms of the metric, the affine connection and the derivatives of the affine connection without involving the derivatives of the metric.

If one wants to think of the variation in terms of the metric and its derivatives instead, defining the affine connection a priori in terms of the metric and its derivatives, as we want to do, then it is more difficult to justify setting the surface variation to zero at the  boundary. As regards the possibility of imposed boundary conditions being incompatible with the equations of motion, note that the momenta have the same number of components as the derivatives of the metric and hence we should be able to leave the metric components as arbitrary and achieve the momenta constraints just by manipulating the derivatives. Thus, that issue seems to have been addressed, at least at first sight. The second issue, related to the uncertainty principle, is not applicable here as we have to fix just the momenta among the canonically conjugate variables.

Thus, we can make  a case for the use of $g_{ab}$ as the dynamical variables as opposed to $g^{ab}$. We shall call the pair $(g_{ab}, \sqg M^{cab})$ as ``holographically conjugate'' variables (HCVs) in the spirit of the $-\partial(pq)$ structure of the surface term in terms of these variables. In the next section, we will see that the same conclusion can be arrived  in an simpler manner using some  scaling arguments. In the subsection after that, these arguments will lead  us to another pair of ``holographically conjugate'' variables with a symmetric, contravariant, two-indexed object taking the place of $g_{ab}$.

\subsection{Holographically Conjugate Variables through Scaling Arguments}
\label{Holoscaling}
In this section, we shall show that the results of the previous section can also be obtained using simple scaling arguments.  There is an alternate method of arriving at \ref{holorel} using scaling arguments (see Project 6.3 in \cite{gravitation}). We shall describe this method below since it elucidates the role played by the use of $g_{ab}$ as the dynamical variable rather than $g^{ab}$.

Consider a Lagrangian $L(q_{A},\partial_i q_{A})$, where $q_A$ is some dynamical variable and $A$ denotes a collection of indices, such that the Lagrangian  is a homogeneous function of degree $\mu$ in $q_A$ and degree $\lambda$ in $\partial_i q_A$. The Euler-Lagrange function (see \ref{Euler-deriv}) obtained from $L$ for the variable $q_{A}$ is  
\begin{equation}
 F^A\equiv \frac{\partial L}{\partial q_A}-\partial_i \left[\frac{\partial L}{\partial(\partial_i q_A)}\right].
 \label{scale-rel}
\end{equation}
Forming the contraction $q_A F^A$ and using Euler's theorem for homogeneous functions, one can easily show that
\begin{equation}
q_A F^A=(\lambda + \mu)L-\partial_i \left[q_A \frac{\partial L}{\partial(\partial_i q_A)}\right] .\label{holgen}
\end{equation}
If we take $L$ to be $\sqrt{-g}L_{\textrm{quad}}$ and $q_A$ to be $g_{ab}$, we would have $\mu=-1$ and $\lambda=+2$. Also, $F^A=-\sqrt{-g}(R^{ab}-\frac{1}{2}g^{ab}R)=-\sqg G^{ab}$ and $q_A F^A=\sqrt{-g} R$. Hence, \ref{scale-rel} becomes
\begin{eqnarray}
\sqrt{-g} R &=& \sqrt{-g}L_{\textrm{quad}} - \partial_c\left(g_{ab}\frac{\partial \sqrt{-g} L_{\rm quad}}{\partial (\partial_c g_{ab})}\right) \label{holgdd}\\
&=&\sqrt{-g}L_{\textrm{quad}} - \partial_c\left(g_{ab}\sqrt{-g}M^{cab}\right). \label{holgddM}
\end{eqnarray}
On the other hand, for the case of $g^{ab}$ as the variable, we have $\mu=-3$, $\lambda=+2$, $F^A=\sqg G^{ab}$ and $q_A F^A=-\sqrt{-g} R$. Thus, we obtain
\begin{equation}
\sqrt{-g} R = \sqrt{-g}L_{\textrm{quad}} + \partial_c\left(g^{ab}\frac{\partial \sqrt{-g} L_{\rm quad}}{\partial (\partial_c g^{ab})}\right) = \sqrt{-g}L_{\textrm{quad}} + \partial_c\left[g^{ab}(-\sqrt{-g}M^{c}_{ab})\right]~,
\label{holguu}
\end{equation}
with a crucial sign difference in the second term.

In both cases, the second derivatives are confined to the surface term. Hence, if the dynamical variables and their derivatives are fixed on the boundary of the volume under consideration, the equations of motion can be obtained by varying $\mathcal{L}_{\rm quad}$ alone and will be of second order in the derivatives of the metric. As has been explained in the introduction, we cannot fix both the dynamical variable and its corresponding momentum at the boundaries.  Now, the surface term that arises from the variation of $\mathcal{L}_{\rm quad}$ alone will be of the form $\partial(p \delta q)$ and can be eliminated by fixing the dynamical variable $q$ at the boundary. But $\mLq$ has the disadvantage that it is not a tensor density and, in fact, vanishes in a local inertial frame. A $-\partial(pq)$ term added to this Lagrangian density, as in the case of $q=g_{ab}$ in \ref{holgdd}, will make it into a tensor density and \textit{also} modify the surface term in the variation to the form $-\partial(q\delta p)$, allowing us to fix just the canonical momenta at the boundary and obtain the equations of motion (see section II in \cite{Mukhopadhyay:2006vu} for a detailed discussion). The explicit variation of \ref{holgddM} is
\begin{equation}
\delta(\sqrt{-g}R)=-\sqrt{-g}G^{ab}\delta g_{ab}-\partial_{c}[g_{ik}\delta(\sqrt{-g} M^{cik})],  \label{varRMg}
\end{equation}
showing that we only need to fix $\sqrt{-g} M^{cik}$ at the boundaries to obtain the equations of motion. This is equivalent to fixing the variation of the connection at the boundaries, as can be seen from \ref{vargMvarfN}. This insight suggests that Hilbert action  is better considered as describing a theory in the space of the canonical momenta corresponding to $g_{ab}$.

On the other hand, \ref{holguu} tells us that the surface term occurs with the wrong sign when the variable $g^{ab}$ is used. The explicit variation in this case leads to
\begin{equation}
\df(\sqg R)=\sqrt{-g}G_{ab}\delta g^{ab}-\partial_{c}[g^{ik}\delta(\sqrt{-g} M^{c}_{ik})]-2\partial_c[\sqg M^{c}_{lm} \df g^{lm}].                  
\end{equation}
The surface term now has the variations of both the dynamical variables and the momenta. Generally, textbooks choose $g^{ab}$ as the dynamical variable because some extra care has to be taken to derive the equations of motion if $g_{ab}$ is taken as the variable (see Exercise 6.9 in \cite{gravitation}). But then one would either have to ignore the variation in the surface term or cancel it with a counter-term like the Gibbons-Hawking-York counter-term \cite{York:1972sj,Gibbons:1976ue}, but neither option is as simple and neat as using $g_{ab}$ as the variable and fixing the canonical momenta at the boundaries. 

In the next section, we present another object which, along with its canonical momentum, forms a pair of HCVs (holographically conjugate variables) and can be used in place of $g_{ab}$ for a variational approach to general relativity.

\subsection{An Alternate Pair of Holographically Conjugate Variables}
\label{subsec:fN}
A natural question that arises  is whether $g_{ab}$ is unique in providing us with a neat and clean variational principle for general relativity. In \ref{app:scaling}, we have used scaling arguments to investigate if there are other variables which may be used in place of $g_{ab}$ in the variational approach to general relativity. Although $g^{ab}$ cannot be considered as a good variable for this purpose, we have found an object with two contravariant indices which appears suitable. This object is the tensor density $\sqg g^{ab}$, denoted in this paper by $f^{ab}$ for typographical convenience. The variable $f^{ab}$ scales linearly with $g_{ab}$ i.e $g_{ab}\to\alpha g_{ab}$ leads to  $f^{ab}\to \alpha f^{ab}$. It is precisely this linear scaling that makes $f^{ab}$ a suitable substitute for $g_{ab}$, as demonstrated in \ref{app:scaling}. We shall use the symbol $f_{ab}$ for the corresponding covariant tensor density, $\sqg g_{ab}$. Note that $f_{ab}$ is not the inverse of $f^{ab}$. See \ref{app:fab} for some useful properties and relations pertaining to $f^{ab}$ and its variation. 

As proved in \ref{app:scaling}, writing the Einstein-Hilbert action in terms of the new variable $f^{ab}$ reproduces the structure of \ref{holgdd}. We obtain 
\begin{eqnarray}
\sqg R =  \sqg \Lq - \partial_{c}[f^{ab}\frac{\partial (\sqg L_{\textrm{quad}} )}{\partial( \partial_c f^{ab})}] =\sqg \Lq - \partial_{c}[f^{ab}N^{c}_{ab}]~, \label{holoNf}
\end{eqnarray}
where we have defined the object $N^{c}_{ik}$ to be the canonical momentum corresponding to $f^{ab}$ for $\mLq$ by:
\begin{equation}
N^{c}_{ik}\equiv \frac{\partial (\sqg L_{\textrm{quad}} )}{\partial( \partial_c f^{ik})}~. \label{Ndef}
\end{equation}
The explicit expression for $N^{c}_{ik}$ has already been written down in \ref{NGamma}. The reason for the demand of symmetrisation in \ref{NGamma} is evident now. Note that it is a simpler object than $M^{cik}$ and is constructed from the affine connection alone.

Next, let us look at the expressions for variation of the Einstein-Hilbert action in terms of $f^{ab}$ and $N^{c}_{ab}$. As we have argued in \ref{Holoscaling}, the $-\partial (qp)$ structure in \ref{holoNf} gives rise to a $-\partial(q \df p)$ surface term, along with the bulk term which provides the equations of motion as usual. Using \ref{varRMg1}, \ref{vargMvarfN} and \ref{GgRf}, we can write down the variation as
\begin{eqnarray}
\delta(\sqrt{-g}R)&=& R_{ab}\delta f^{ab}+ f^{ab} \df R_{ab} = R_{ab}\delta f^{ab}-\partial_{c}[f^{ik}\delta N^{c}_{ik}], \label{varRNf}
\end{eqnarray}

In addition to their utility in simplifying the variation of the action, there are two main advantages to the HCVs (holographically conjugate variable pairs) $(g_{ab}, \sqg M^{cab})$ and $(f^{ab}, N^{c}_{ab})$. The first one is that many known expressions and formulae simplify considerably when written in terms of these variables and make our theoretical life  easier. The pair of variables $(f^{ab}, N^{c}_{ab})$  gives a particularly stellar performance in this regard and will be our variables of choice for most of the work in this paper. More importantly, the \textit{variations of these variables on a horizon will be shown to have a direct thermodynamical interpretation}. In the next section, we shall demonstrate the first point by writing down several well-known expressions and formulae in terms of the HCVs (holographically conjugate variables pairs).

Historically, it was indeed noted in the early days of general relativity that $f^{ab}$ is a good variable to use and was even given preference over $g_{ab}$ at times \cite{Eddington:1924,Schrod:1950}. The variables $f^{ab},N^{c}_{ab}$ were later used, albeit the non-symmetrised version of $N^{c}_{ab}$, by Einstein in his work with Kauffman attempting to go beyond general relativity \cite{Einstein-Kauffman,meaning}. A recent paper that uses these variables but does not make the identification of $N^{c}_{ab}$ as the canonical momentum corresponding to $f^{ab}$ is \cite{Kijowski:97}. The paper \cite{Babak:1999dc} contains the analogues of some of the expressions in terms of these variables given in the next section, but the calculations have been done as perturbations on a flat metric. We believe the holographic relation between the bulk and the surface terms of the action gives a simple and elegant motivation for using the variables $(f^{ab}, N^{c}_{ab})$.

%#######################################################################
\section{Gravity in terms of $f^{ab}$ and $N^{c}_{ab}$}
\label{canonvar}
In this section, we shall show that many objects of common use have simpler expressions in terms of $f^{ab}$ and $N^{c}_{ab}$ than in the conventional description. First, note that \ref{dfwithdg} allows us to write
\begin{equation}
\partial_c f^{ab}=\sqg B^{ab}_{lm} \partial_c g^{lm}; \quad
B^{lm}_{ab}\equiv \frac{1}{2}(\df^{l}_{a}\df^{m}_{b} + \df^{l}_{b}\df^{m}_{a}) -(1/2)g^{lm}g_{ab}
\label{derfinderg}
\end{equation}
and since $B^{ab}_{lm}$ is independent of the derivatives of $g^{ab}$, we can write
\begin{equation}
\frac{\partial(\partial_c f^{ab})}{\partial(\partial_d g^{lm})}= \df^{d}_{c} \sqg B^{ab}_{lm} .
\end{equation}
Using this relation and \ref{Mudd} and \ref{Ndef}, we get the relation
\begin{equation}
\sqg M^{c}_{ab}=-\sqg B^{lm}_{ab}N^{c}_{lm},
\end{equation}
which enables us to relate the canonical momenta corresponding to $g_{ab}$ and $f^{ab}$ by:
\begin{equation}
M^{cab} = -B^{ab}_{de}N^{cde}. 
\label{MwithN}
\end{equation}
This can be easily inverted using \ref{Bnorm} to give
\begin{eqnarray}
N^{cab} = -B^{ab}_{de}M^{cde} 
        =-(M^{cab}+\frac{1}{2}g^{ab}V^{c}) .
\label{NinM}	
\end{eqnarray}
Looking at \ref{M_in_GammaV}, we can see that $N^{a}_{bc}$ is related to the Christoffel symbols by the simple expression \ref{NGamma}.
Thus, we can infer that although $N^{a}_{bc}$ is not a tensor, $\df N^{c}_{ab}$ is a tensor because $\df \Gamma^{c}_{ab}$ is a tensor.
\ref{NinM} can be readily converted to an expression in terms of the first derivatives of the metric using \ref{M3}. This expression is
\begin{equation}
N^{cab} = -\frac{1}{2}B^{ab}_{de}M^{cdeijk}\partial_i g_{jk}. 
\end{equation}
We can also write $N^{c}_{ab}$ in terms of $\partial_c f^{ab}$ using the inverse of \ref{derfinderg}, to obtain:
\begin{equation}
N^{c}_{ab}=\frac{-1}{2 \sqg}\left[g^{ci}B_{abrs}-\df^{c}_{r}(\df^{i}_{a}g_{bs}+\df^{i}_{b}g_{as}) \right]\partial_i f^{rs} 
\end{equation}
To replace the first derivatives in the theory with canonical momenta, we need to invert \ref{NGamma}. This is easily done by assuming the  following ansatz for the Christoffel symbols:
\begin{equation}
\Gamma^{c}_{ab}=a N^{c}_{ab}+b(N^{d}_{ad}\delta^{c}_{b}+N^{d}_{bd}\delta^{c}_{a}).
\end{equation}
and
substituting back in \ref{NGamma} to solve for $a$ and $b$. We obtain
\begin{equation}
\Gamma^{c}_{ab}= -N^{c}_{ab}+\frac{1}{3}(N^{d}_{ad}\delta^{c}_{b}+N^{d}_{bd}\delta^{c}_{a}). \label{Gamma_in_N}
\end{equation}
Now we can substitute for $N^{c}_{ab}$ in terms of $M^{c}_{ab}$. (We shall not display this equation explicitly as it does not appear to simplify further.)
We can use the relation $\partial_c g_{ab}= \Gamma_{a,bc}+\{a\leftrightarrow b\}$ and obtain
the derivatives of the metric to be:
\begin{equation}
\partial_c g_{ab}= \{-g_{ae}[N^{e}_{bc}-\frac{1}{3}(N^{d}_{cd}\delta^{e}_{b}+N^{d}_{bd}\delta^{e}_{c})]\} + \{a\leftrightarrow b\}, 
\end{equation}
and, further
\begin{equation}
\partial_c f^{ab} = [f^{ad}(N^{b}_{dc}-\frac{1}{3}\df^{b}_{c}N^{e}_{de})]+[a\leftrightarrow b]~. \label{dfinN} 
\end{equation}
We shall now make use of these expressions to express the Lagrangian, curvature tensor etc. in terms of the HCVs (holographically conjugate variables).

\subsection{Riemann Tensor, Ricci Tensor and Ricci Scalar}

We next give the formulas for the Riemann tensor, the Ricci tensor and the Ricci scalar in terms of $N^{a}_{bc}$ for ready reference:
\begin{eqnarray}
R^{a}_{bcd}&=& [-(\df^{e}_{b}\partial_c+N^{e}_{bc})(N^{a}_{de}-\frac{1}{3}\df^{a}_{d}N^{f}_{ef})+\frac{1}{9}\df^{a}_{c}N^{e}_{be}N^{f}_{df}]-[c\leftrightarrow d]   \\
R_{ab}&=&-(\partial_c N^{c}_{ab}+N^{c}_{ad}N^{d}_{bc}- \frac{1}{3} N^{c}_{ac}N^{d}_{bd})  \label{RabN}  \\
R &=& -g^{ab}\partial_c N^{c}_{ab}-\Lq  \\
\sqg R &=& -f^{ab}\partial_c N^{c}_{ab}-\frac{1}{2}N^{c}_{ab}\partial_c f^{ab} \\
&=&  \frac{1}{2}N^{c}_{ab}\partial_c f^{ab}- \partial_c (f^{ab}N^{c}_{ab})\label{holfN1}\\
&=&-\frac{1}{2}[f^{ab}\partial_c N^{c}_{ab} +\partial_c (f^{ab}N^{c}_{ab})]\label{holfN2}
\end{eqnarray}
Here, \ref{holfN1} is the usual decomposition of $\sqg R$ into a bulk term and a surface term as given in \ref{decomp} while \ref{holfN2} is an alternate decomposition in which the bulk term contains second derivatives of the metric.

\subsection{The Lagrangian}

We can substitute for $\Gamma$s in the bulk term in \ref{bulksur} in terms of $N^{c}_{ab}$s from \ref{Gamma_in_N} and obtain the bulk part of the Lagrangian to be:
\begin{eqnarray}
L_{\rm{quad}} = g^{bd}N^{i}_{dj}N^{k}_{bl}[\delta^{l}_i\delta^{j}_k-\frac{1}{3}\delta^{j}_i\delta^{l}_k] = g^{bd} \left(Tr[N_b N_d]-\frac{1}{3}Tr[N_b]Tr[N_d]\right)   
\label{quad_in_N}
\end{eqnarray}
where we have taken $N^{c}_{ab}$ to be the $cb\textrm{th}$ element of a matrix $N_a$, and $Tr[N_a]$ is the trace of this matrix. This also allows us to write:
\begin{equation}
\mLq = \frac{1}{2}N^{c}_{ab}\partial_{c}f^{ab}~,
\label{quad_in_N1}
\end{equation} 
with striking simplicity.
Note that this exhibits a `$p\dot{q}/2$' structure, which is a consequence  of the fact that the Lagrangian is quadratic in $\dot{q}$. We can also write $L_{quad}$ in terms of $M^{c}_{ab}$ as
\begin{eqnarray}
L_{\rm{quad}}&=&g^{bd}M^{c}_{dk}M^{k}_{bc}-\frac{2}{3}g^{ir}M^{k}_{ir}M^{d}_{dk}-\frac{1}{3}g^{bk}M^{i}_{bi}M^{d}_{dk}+\frac{1}{6}g_{dk}g^{ir}M^{d}_{ir}g^{xy}M^{k}_{xy} \nonumber\\
&=&g^{bd}Tr[M_b M_d]+\frac{2}{3}Tr[M_k]V^{k}-\frac{1}{3}g^{bd}Tr[M_b]Tr[M_d]+\frac{1}{6}g_{bd}V^{b}V^{d}~,
\end{eqnarray}
which, in comparison to \ref{quad_in_N}, appears to be quite complicated. But, from \ref{derivLq} and \ref{M3}, the `$p\dot{q}/2$' structure arises with the variables $(g_{ab},\sqg M^{cab})$ also, as given below:
\begin{equation}
\mLq = \frac{\sqg}{2}M^{cab}\partial_{c}g_{ab}~. 
\end{equation}
The formulas for $\Ls$ have already been given in \ref{surVc} and \ref{holoNf}. We have
\begin{eqnarray}
\mLs=\partial_{c}(\sqg V^{c}) = -\partial_c[g_{ab}(\sqg M^{cab})]
= -\partial_c[f^{ab}N^{c}_{ab})].
\end{eqnarray}
In fact, the above relations hold even without the derivatives,
\begin{eqnarray}
\sqg V^{c} = -g_{ab}(\sqg M^{cab}) \label{VcM}
= -f^{ab}N^{c}_{ab}, \label{VcN}
\end{eqnarray}
This equation will be of use in the study of horizon thermodynamics.
%##############################################################################################
\subsection{Euler Derivative and the Equations of Motion}
\label{Euler-deriv}
The Euler derivative of any function $K[\phi, \partial_i \phi, ...]$ with respect to the variable $\phi$ is defined as
\begin{equation}
\frac{\df K[\phi, \partial_i \phi, ...]}{\df \phi} = \frac{\partial K[\phi, \partial_i \phi, ...]}{\partial \phi} - \partial_a\left[\frac{\partial K[\phi, \partial_i \phi, ...]}{\partial( \partial_a \phi)} \right] + \partial_a \partial_b \left[\frac{\partial K[\phi, \partial_i \phi, ...]}{\partial( \partial_a \partial_b \phi)} \right]- ... 
\end{equation}
The function thus obtained is also called the Euler-Lagrange function resulting from $K$ for the variable $\phi$. We will use the notation $E[K, \phi]$ for this object. We have already made use of the Euler-Lagrange functions in \ref{Holoscaling}.

The most general variation of $K$ can be written as
\begin{equation}
\df K = \frac{\df K}{\df \phi} + \textrm{ surface variations}
\end{equation}
Thus, when the surface terms can be consistently put to zero, the equations of motion are obtained by equating the Euler derivative to zero (which is the origin of the nomenclature).
We shall now list the Euler-Lagrange functions obtained from $\mLq$ for the dynamical variables under our consideration, namely $g^{ab}$, $g_{ab}$ and $f^{ab}$.  The functions are, respectively,
\begin{align}
E[\mLq,g^{ab}] &= \sqg G_{ab}~, \\
E[\mLq,g_{ab}] &= -\sqg G^{ab}\textrm{ and}\\
E[\mLq,f^{ab}] &= R_{ab}~. 
\end{align}
The equations of motion in the absence of matter can be obtained by equating these functions to zero. 
In the presence of matter, the matter Lagrangian has to be added to $\mLq$ and the Euler-Lagrange function of the total Lagrangian has to be obtained. We then get the equivalent expressions: 
\begin{equation}
G_{ab}=8 \pi T_{ab};\quad G^{ab}= 8 \pi T_{ab};\quad R_{ab}=8 \pi (T_{ab} - \frac{g_{ab}}{2} T)~. 
\end{equation}
From \ref{RabN}, we can write down the equations of motion as
\begin{eqnarray}
\partial_c N^{c}_{ab}&=& -N^{c}_{ad}N^{d}_{bc}+ \frac{1}{3} N^{c}_{ac}N^{d}_{bd}-8 \pi (T_{ab} - \frac{g_{ab}}{2} T)~,
\end{eqnarray}
a first order differential equation in $N^{c}_{ab}$ with a matter source term.  
In the local inertial frame, the equation becomes
\begin{equation}
\partial_c N^{c}_{ab}=-8 \pi (T_{ab} - \frac{\eta_{ab}}{2} T)
\end{equation}
which gives a conservation law for the canonical momenta $N^{c}_{ab}$ valid in a space-time volume small enough to respect our local inertial frame.
The corresponding equations of motion in terms of $M^{cab}$ look even nicer in the local inertial frame. They are $G^{ab}=\partial_c M^{cab}= 8 \pi T^{ab}$ i.e. 
\begin{eqnarray}
 \partial_c M^{cab}= 8 \pi T^{ab}
\end{eqnarray}
Note that in the local inertial frame $\sqg$ is unity and hence $M^{cab}=\sqg M^{cab}$ plays the role of canonical momentum corresponding to the metric.   
Thus, if we take a region small enough for a local inertial frame to be enforced, the surface integral of the canonical momenta corresponding to the metric components gives the components of the matter energy-momentum tensor contained in that volume. Such a balance between gravitational variables and matter variables will again arise when we study horizon thermodynamics.
%##############################################################################################################
\subsection{Palatini Variational Principle and Hamilton's Equations for Gravity} 
\label{sec:Palatini}
In classical mechanics, the variation of the action is carried out regarding the dynamical variable $q$ and its first time-derivative, $\dot{q}$ as the independent variables, for the class of actions which do not depend on the higher time-derivatives of $q$. The momentum $p$ is then defined as the partial derivative of the Lagrangian with respect to $\dot{q}$. There is an alternate way of carrying out the variation considering $q$ and $p$ as independent variables, with the Lagrangian being defined as $L=p \dot{q} - H(q,p)$, where $H(q,p)$ is the Hamiltonian of the system \cite{MTW, Goldstein}. In this case, after fixing the variations of $q$ at the boundary, the variation with respect to $p$ gives the equation $\dot{q}= \partial H/\partial p$, while the variation with respect to $q$ gives $\dot{p}= -\partial H/ \partial q$, which are the well-known Hamilton's equations. This variational principle is referred to as modified Hamilton's principle.

Analogously there are two well-known variational principles in general relativity: one in which the components of the affine connection are considered as given in terms of the derivatives of the metric and the variation of the Einstein-Hilbert action is carried out in terms of the variation of the metric and its first and second derivatives; and one in which the metric and the affine connection are considered as independent and varied separately. The second method is called the Palatini variational principle \cite{Palatini}. But, as we have been arguing the case for the use of the HCVs (holographically conjugate variables), we shall now outline a version of the Palatini variational principle in terms of the variables $(f^{ab}, N^{c}_{ab})$. As these form a $(q,p)$ pair, this would be a direct analogue of our classical mechanics exposition of the alternate variational principle. This aspect {\it is in contrast with the usual approach} in which the metric and connection are treated as independent variables, because the connection $\Gamma^a_{bc}$ is not the canonically conjugate variable to the metric $g_{bc}$.

Using \ref{RabN}, the Einstein-Hilbert Lagrangian can be expressed as
\begin{eqnarray}
\sqrt{-g}R = f^{ab}R_{ab}= f^{ab}(-\partial_c N^{c}_{ab}-N^{c}_{ad}N^{d}_{bc}+ \frac{1}{3} N^{c}_{ac}N^{d}_{bd})~.
\label{palatini1}
\end{eqnarray}
We will now vary this Lagrangian considering $f^{ab}$ and $N^i_{ab}$ as independent variables. Note that $R_{ab}$ is a function only of $N^i_{ab}$ and is independent of $f^{ab}$, which should not be surprising as we know $R_{ab}$ to be a function only of the affine connection. The variation of the action (\ref{palatini1}) with respect to $N^i_{ab}$ is given by
\begin{eqnarray}
\delta\Big(\sqrt{-g}R\Big)|_{f^{ab}} &=& f^{ab} \df R_{ab} \\
&=& \Big[\partial_c f^{ab} - 2f^{ad}N^b_{cd} + \frac{2}{3}f^{am}N^d_{dm}\delta^b_c\Big]\delta N^c_{ab} 
- \partial_c (f^{ab} \df N^{c}_{ab}) ~.
\label{palatini2}
\end{eqnarray}
Once  we fix $N^{a}_{bc}$ at the boundary, we obtain the corresponding equations of motion by equating the symmetrised coefficient of $\df N^{c}_{ab}$  to zero. These equations are
\begin{eqnarray}
 \partial_c f^{ab}&=& f^{ad}N^b_{cd} + f^{bd}N^a_{cd} - \frac{1}{3}f^{am}N^d_{dm}\delta^b_c -  \frac{1}{3}f^{bm}N^d_{dm}\delta^a_c \label{dfinN1}
\end{eqnarray}
Note that \ref{dfinN1} is identical with \ref{dfinN}. Thus, the action principle dictates the connection between $N^{c}_{ab}$ and $(f^{ab},\partial_c f^{ab})$.
In order to connect this up with the standard result $\nabla_c g^{ab}=0$ obtained from Palatini variational principle in its conventional form, we need to keep in mind that the covariant derivative is, as of now, defined as usual in terms of the affine connection but the affine connection is not related to the metric or its derivatives. Substituting $f^{ab}= \sqg g^{ab}$ in \ref{dfinN1} and contracting both sides with $g_{ab}$, we can obtain the relation
$N^{d}_{cd}= (3/4) g^{ab} \partial_c g_{ab}$ which gives us:
\begin{equation}
\sqg N^{d}_{cd}= \frac{3}{2}\partial_c \sqg; \,\,\  i.e., \,\,\ \sqg \Gamma^{d}_{cd}= \partial_c \sqg  \label{GammaNinsqg}
\end{equation}
We shall now use this result to evaluate $\sqrt{-g}\nabla_cg^{ab}$. Expanding $\sqrt{-g}\nabla_cg^{ab}$, we find
\begin{eqnarray}
\sqrt{-g}\nabla_cg^{ab} = \partial_c f^{ab} - g^{ab}\partial_c\sqrt{-g} + f^{mb}\Gamma^a_{cm} + f^{ma}\Gamma^b_{cm}~.
\label{palatini7}
\end{eqnarray}
Next, using the identity $\partial_c\sqrt{-g} = (2/3)\sqrt{-g}N^d_{cd}$ from \ref{GammaNinsqg} and expressing the connections in terms of $N^i_{ab}$ variables by \ref{NGamma}, we obtain
\begin{eqnarray}
\sqrt{-g}\nabla_cg^{ab} = \partial_c f^{ab}  - f^{ad}N^b_{cd} - f^{bd}N^a_{cd} + \frac{1}{3} f^{am}N^d_{md}\delta^b_c + \frac{1}{3} f^{bm}N^d_{md}\delta^a_c~,
\label{palatini9}
\end{eqnarray}
which vanishes due to \ref{dfinN1}.
Hence, the equation of motion (\ref{dfinN1}) is precisely the metricity condition obtained by the variation of the connection in conventional Palatini formalism.

We shall now provide an alternate perspective on these results which would prove quite fruitful. Note that \ref{dfinN1} can be written as
\begin{equation}
\partial_c f^{ab}= -\frac{\partial}{\partial N^{c}_{ab}} (\sqg R+ f^{ab}\partial_c N^{c}_{ab})\label{dfinRNf}
\end{equation}
We have obtained here an analogue of the Hamilton's equation $\dot{q}= \partial H/\partial p$. The analogy can be made more precise as follows. Let us define:
\begin{equation}
\mathcal{H}_g=f^{ab}(N^{c}_{ad}N^{d}_{bc}- \frac{1}{3} N^{c}_{ac}N^{d}_{bd}) \label{Hg}
\end{equation}
Then, with the notional correspondence $f^{ab}\longrightarrow q$ and $N^{c}_{ab}\longrightarrow p$, we can establish:
\begin{equation}
\sqg R \longrightarrow  -q \partial p-\mathcal{H}_g = \{p \partial q - \mathcal{H}_g\}-\partial(qp) = \mathcal{L_{\textrm{quad}}+L_{\textrm{sur}}}
\end{equation}
Comparing \ref{Hg} with \ref{quad_in_N}, we see that
\begin{equation}
\mathcal{H}_g = \mLq \longrightarrow \frac{1}{2}p \partial q
\label{Lquad}
\end{equation}
Thus, the quadratic Lagrangian density that is used to derive the equations of motion can be also be interpreted as a Hamiltonian density. The equation \ref{dfinRNf} 
can then be rewritten in the desired form as
\begin{equation}
\partial_c  f^{ab} = \frac{\partial \mathcal{H}_g}{\partial N^{c}_{ab}}
\end{equation}

Proceeding by analogy, our next stop would be the other Hamilton's equation of motion, $\dot{p}= -\partial H/ \partial q$. Consider the variation of the action (\ref{palatini1}) with respect to the $q$ variable $f^{ab}$, given by
\begin{eqnarray}
\delta\Big(\sqrt{-g}R\Big)|_{N^i_{ab}} = R_{ab}\delta f^{ab}~.
\label{palatini4}
\end{eqnarray}
The equation of motion obtained on extremising the action with respect to variations in $f^{ab}$ is $R_{ab}=0$ which is the same as:
\begin{eqnarray}
  -N^{c}_{ad}N^{d}_{bc}+ \frac{1}{3} N^{c}_{ac}N^{d}_{bd} =\partial_c N^{c}_{ab} 
\end{eqnarray}
Referring back to \ref{Hg}, we see that this equation can be expressed as
\begin{equation}
\partial_c N^{c}_{ab}=-\frac{\partial \mathcal{H}_g}{\partial f^{ab}} \label{Hamilton2}
\end{equation}
giving the second of the Hamilton's equations. But unlike the case in classical mechanics where the momentum would be conserved in the absence of external forces, we see that $N^{c}_{ab}$ is capable of driving its own change, due to the nonlinear nature of gravity. 

The next natural step is to consider the inclusion of the matter Lagrangian density $\mLm=\sqg L_{\textrm{m}}$. This can be accomplished by defining a total Hamiltonian as
\begin{equation}
\mH_{\textrm{tot}} = \mH_g - \mLm
\end{equation}
If the first term $\mH_g$, which is equal to $\mLq$, be considered as a kinetic term, then it is natural to think of $\mLm$ as a potential term as far as gravity is concerned. We shall make here the assumption that the $\mLm$ under consideration depends only on $f^{ab}$ and not on $N^{c}_{ab}$. In such a case, \ref{dfinN1} retains its form. Thus, if we choose to express everything in terms of the metric and its derivative, our assumption is tantamount to the assumption that $\mLm$ does not depend on the derivatives of the metric. (This is similar to the case in classical mechanics when we consider velocity-independent potentials.)

If we now we take the usual definition of the matter energy-momentum tensor as
\begin{eqnarray}
T_{ab} = -\frac{2}{\sqrt{-g}}\frac{\partial(\sqrt{-g}L_m)}{\partial g^{ab}}~,
\label{palatini10}
\end{eqnarray}
we can obtain the following equality:
\begin{equation}
\frac{\partial \mLm}{\partial f^{ab}}= -\frac{1}{2}\Big[ T_{ab} - \frac{1}{2} T g_{ab}\Big]= - \frac{1}{2} B^{lm}_{ab}T_{lm}\equiv- \frac{1}{2} \overline{T}_{ab}~, 
\end{equation}
where $B^{lm}_{ab}$ was defined in \ref{app:fab}. We have defined a new object $\overline{T}_{ab}$ here, which bears to $T_{ab}$ the same relation as $G_{ab}$ bears to $R_{ab}$.

\ref{Hamilton2} is now modified to read
\begin{eqnarray}
\partial_c N^{c}_{ab}= -\frac{\partial \mathcal{H}_g}{\partial f^{ab}}+\frac{\partial (16\pi \mLm)}{\partial f^{ab}}= -N^{c}_{ad}N^{d}_{bc}+ \frac{1}{3} N^{c}_{ac}N^{d}_{bd}-8\pi \overline{T}_{ab}  \label{EOM_N+m}
\end{eqnarray}
Thus, the presence of matter introduces an extra source term in the equations governing the evolution of $N^{c}_{ab}$. Using \ref{RabN}, it is easy to verify that \ref{EOM_N+m} is equivalent to the usual Einstein's equation of motion, $R_{ab} = 8\pi G \Big(T_{ab} - (1/2) T g_{ab}\Big)$. 

Finally,
note that the variation of the total Lagrangian, gravitational plus matter, on varying $f^{ab}$ and $N^{c}_{ab}$ is given by
\begin{eqnarray}
\df(\sqg R+ 16\pi \mLm) &=&\Big[\partial_c f^{ab} - f^{ad}N^b_{cd} + \frac{1}{3}f^{am}N^d_{dm}\delta^b_c\Big]\delta N^c_{ab} -[(\partial_c N^{c}_{ab}+N^{c}_{ad}N^{d}_{bc}- \frac{1}{3} N^{c}_{ac}N^{d}_{bd})+8\pi \overline{T}_{ab}] \df f^{ab} 
\nonumber
\\
&&- \partial_c (f^{ab} \df N^{c}_{ab})
\end{eqnarray}
The surface variation given in the second line of the above equation contains only variations of $N^{c}_{ab}$ and not of $f^{ab}$, which happened essentially because the Lagrangian with which we started with does not have derivatives of $f^{ab}$. Thus, we need to fix only the ``momenta'' $N^{c}_{ab}$ at the boundary to obtain the equations of motion.

%##############################################################################################################
\subsection{Noether Current}
 So far, we have introduced two sets of naturally conjugate variables and expressed some relevant quantities in general relativity in terms of them. The usual Einstein's equations have also been derived from the variation of these variables.  
 In this subsection, we shall relate the Noether current --- which arises from the diffeomorphism invariance of the action --- with the variations of $N^a_{bc}$ and $f^{ab}$. For a general covariant Lagrangian $L(g_{ab},R_{abcd})$, the Noether current, corresponding to the diffeomorphism $x^a\rightarrow x^a+\xi^a$, can be shown to be related to the Lie derivative of $\Gamma^a_{bc}$ with a particular contraction of the indices with the quantity $P^{abcd} \equiv \partial L/\partial R_{abcd}$. Since $N^a_{bc}$ is a linear combination of the connections, we can easily convert this relation into the required relation.
 
 To do this, we first recall the general form of the Noether current for a Lagrangian $L(g_{ab},R_{abcd})$, corresponding to the diffeomorphism $x^a\rightarrow x^a+\xi^a$. It is given by (see, e.g., Project $8.1$ of \cite{gravitation}).
\begin{eqnarray}
J^a = 2E^a_b\xi^b + L\xi^a + \delta_{\xi} v^a~,
\label{lie6}
\end{eqnarray}
where $\delta_{\xi} v^a$ is such that $\nabla_a(\delta_{\xi} v^a) $ is the  surface term in the variation of the Lagrangian under the diffeomorphism and $E^a_{b} = 0$ is the equation of motion.  For the Lanczos-Lovelock models of gravity, for example, $E^a_{b} = P^{akij}R_{bkij} - \frac{1}{2}\delta^a_{b}L$ and 
$\delta_{\xi} v^j = 2 P^{ibjd}\nabla_b\delta_{\xi}g_{di}$ (see Section 15.4 in \cite{gravitation}), 
where $P^{abcd} \equiv(\partial L/\partial R_{abcd})$ as already mentioned. The tensor $P^{abcd}$ has the algebraic properties of the curvature tensor and, additionally, it is assumed to be divergence-free in Lanczos-Lovelock models:  $\nabla_a P^{abcd} = 0$.
Hence, for Lanczos-Lovelock models, the relation between the surface contribution of the variation of action and the Noether current is 
\begin{equation}
\delta_\xi v^a = J^a - 2{\cal{R}}^a_b\xi^b~, \label{xi_J}
\end{equation} 
where ${\cal{R}}^a_b \equiv  P^{akij} R_{bkij}$. 
We now switch to the language of Lie derivatives. For an indexed object $A$, while $\df_{\xi} A\equiv A'(x)-A(x)$ is the functional change in $A$ at the same value of the space-time coordinates under an infinitesimal coordinate transformation from $x^i$ to $x^i + \xi^i$, the Lie derivative of $A$ is defined as $\pounds_{\xi}A=A(x)-A'(x)$, thus giving $\pounds_{\xi}=-\delta_{\xi}$. Now, we can use $\pounds_{\xi} g_{di} = \nabla_d\xi_i + \nabla_i\xi_d$ to find 
\begin{eqnarray}
\pounds_{\xi} v^a = -2 P^{abdi}\nabla_b\nabla_d\xi_i - 4 P^{aibd}\nabla_b\nabla_d\xi_i =  -2 P^{abdi}\nabla_b\nabla_d\xi_i + 2 R^{a}_{m}\xi^m~,
\label{delta v1}
\end{eqnarray}
where,  in the last step, we have used  $P^{aibd}\nabla_b\nabla_d\xi_i = - (1/2) P^{aibd}R_{mibd}\xi^m$. Therefore, the Noether current turns out to be
\begin{eqnarray}
J^a = 2{\cal{R}}^a_b\xi^b + \delta_\xi v^a = 2{\cal{R}}^a_b\xi^b - \pounds_\xi v^a= 2  P^{abdi}\nabla_b\nabla_d\xi_i~,
\label{lie9}
\end{eqnarray}

We will now re-express this in terms of the variation of the connection. We know that, although $\Gamma^a_{bc}$ is not a tensor, its Lie derivative is a  tensor:
\begin{eqnarray}
\pounds_{\xi}\Gamma^a_{bc} = \nabla_b\nabla_c\xi^a + R^a_{\phantom{a}cmb}\xi^m~.
\label{lie1}
\end{eqnarray}
The Lie variation of $N^i_{ab}$ is likewise a tensor. Using its definition \ref{NGamma} in terms of connection and \ref{lie1}, we obtain the explicit expression:
\begin{eqnarray}
\pounds_\xi N^i_{ab} = - \nabla_a\nabla_b\xi^i + \frac{1}{2}\Big(\delta^i_a\nabla_b\nabla_m\xi^m + \delta^i_b\nabla_a\nabla_m\xi^m\Big) - R^i_{bma}\xi^m~.
\label{lie4}
\end{eqnarray}
Now, contracting \ref{lie1} with $P^{ibc}_ {\phantom{ibc}a}$ we find
\begin{eqnarray}
2P^{ibc}_ {\phantom{ibc}a} \pounds_{\xi}\Gamma^a_{bc} = J^i - 2{\cal{R}}^i_m\xi^m~.
\label{lie2}
\end{eqnarray}
Therefore, from \ref{xi_J} and \ref{lie2},
\begin{equation}
\delta_\xi v^i = 2P^{ibc}_ {\phantom{ibc}a} \pounds_{\xi}\Gamma^a_{bc}~. 
\end{equation}
It turns out that, in general relativity, we can easily write this expression in terms of $N^a_{bc}$. To do this, note that in the case of general relativity,
$L=R$ and $P^{ibca} = \frac{1}{2} (g^{ic}g^{ab}-g^{ia}g^{bc})=Q^{ibca}$. Now, since $N^a_{bc}$ is related to $\Gamma^a_{bc}$ by \ref{NGamma}, it is easy to show that
\begin{eqnarray}
2P^{ibc}_ {\phantom{ibc}a} \pounds_{\xi}\Gamma^a_{bc}=g^{xy}\pounds_\xi N^i_{xy} ~.
\label{lie5}
\end{eqnarray}
So, the surface term in the variation of the Lagrangian density $\sqg R$ can be expressed in different ways as (compare with \ref{vargMvarfN})
\begin{equation}
 \sqg \nabla_i (\delta_\xi v^i) =2 \sqg \nabla_i( P^{ibc}_ {\phantom{ibc}a} \pounds_{\xi}\Gamma^a_{bc})=\sqg \nabla_i(g^{xy}\pounds_\xi N^i_{xy})~,
\end{equation}
while the sought-after expression for the Noether current in terms of the variation of $N^{a}_{bc}$ is given by
\begin{equation}
J^{i}= 2{\cal{R}}^i_m\xi^m + g^{xy}\pounds_\xi N^i_{xy} 
\end{equation}
%##############################################################################################################
\subsection{Canonical Energy-Momentum Pseudotensor}

From the bulk part of the Lagrangian, one can define the canonical energy-momentum pseudotensor (up to overall factors, also known as the Einstein pseudotensor \cite{AE,Eddington:1924,Schrod:1950,Dirac:1975}) as:
\begin{eqnarray}
t^{i}_{k} &=& \frac{\partial(\sqg L_{quad})}{\partial(\partial_i g_{ab})}\partial_{k}g_{ab}-\df^{i}_{k} \sqg L_{quad}  \label{tikg}\\
&=&   \frac{\partial(\sqg L_{quad})}{\partial(\partial_i f^{ab})}\partial_{k}f^{ab}- \df^{i}_{k} \sqrt{-f} L_{quad}\label{tikf}\\
&=&   N^{i}_{ab}\partial_{k}f^{ab}- \frac{1}{2}\df^{i}_{k} (N^{c}_{ab}\partial_c f^{ab}),
\label{t}
\end{eqnarray}
which, when taken together with the matter energy-momentum tensor, satisfies the conservation law (see \ref{app:conserv} for a proof): 
\begin{equation}
\partial_{i}(t^{i}_{k}-16 \pi\sqg T^{i}_{k})=0.
\end{equation}
Finally, we mention that the ``trace'' of the pseudotensor can be related to the Hamiltonian ${\cal{H}}_g$, defined by \ref{Hg}. From \ref{t}, after contracting with $\delta_i^k$, we find $t^i_i = -N^c_{ab}\partial_cf^{ab}$ which is $-2{\cal{L}}_{\textrm{quad}}$ (see, \ref{quad_in_N1}). Therefore, using \ref{Lquad},  
\begin{equation}
t^i_i = -2 {\cal{H}}_g~. 
\end{equation}
This concludes our discussion of standard features of general theory of relativity in terms of the canonically
conjugate variables $(f^{ab},N^c_{ab})$. We shall now discuss the connection between the HCVs (holographically conjugate variables) and the thermodynamics of horizons. 
%###############################################################################
\section{Thermodynamics with the Holographically Conjugate Variables}
\label{therm}
In this section, we shall show that the variations $p \df q$ and $q \df p$  obtained from our sets of HCVs (holographically conjugate variables) have direct thermodynamical interpretations when integrated over horizons. We shall first prove the results in a general static spacetime. Then, we shall specialize to the spherically symmetric case and examine Schwarzchild and Reissner-N\"{o}rdstrom horizons in order to obtain a physical feel of our results. Finally, we shall show how our results generalize to integrals over any null surface which acts as a local Rindler horizon.
\subsection{Preliminaries}
We shall first set up a couple of relations involving the two canonical momenta  before we proceed to the thermodynamic relations.
From \ref{varRMg}, \ref{varRNf} and \ref{GgRf}, we see that
\begin{equation}
\partial_{c}[g_{ik}\delta(\sqrt{-g} M^{cik})]                                             
=\partial_{c}[\sqrt{-g}g^{bk}\delta N^{c}_{bk}]. \label{dq_p}
\end{equation}
Since $V^{c}=-g_{ab}(\sqg M^{cab})=-f^{ab}N^{c}_{ab}$, we also have
\begin{equation}
\delta g_{ik}(\sqrt{-g} M^{cik})=\delta f^{ab}N^{c}_{ab} . \label{p_dq}
\end{equation}
The variations $\delta(\sqrt{-g} M^{cxy})$ and $g^{bx}g^{ky}\delta N^{c}_{kb}$ are not equal but become equal on contraction with $g_{xy}$. 
To characterize the difference, we define a tensorial quantity
\begin{equation}
H^{ab,cd}\equiv g^{ac}g^{bd}-(1/4)g^{ab}g^{cd}
\label{H}
\end{equation}
which has the property that $H^{ab,cd}g_{ab}=H^{ab,cd}g_{cd}=0$. This quantity is also a projector since we have $H^{ab}_{cd}H^{cd,ef}=H^{ab,ef}$. In fact, any indexed quantity $Q^{ab..}$ can be written as
\begin{equation}
Q^{ab...}=\frac{1}{4}g^{ab}g_{cd}Q^{cd...}+H^{ab}_{cd}Q^{cd...} .
\label{Hpro}
\end{equation}
where dots indicate indices which are not displayed.
A contraction with $g_{ab}$ will pick up a contribution from the first term and a contraction with $H^{ef}_{ab}$ will catch the second term.
Using $H^{ab,cd}$, we can write
\begin{equation}
\delta(\sqrt{-g} M^{cxy})-\sqg g^{bx}g^{ky}\delta N^{c}_{kb}= 2(H^{ik,xy}\delta^{c}_{z}-H^{ik,cy}\delta^{x}_{z})\delta \Gamma^{z}_{xy},
\end{equation}
which makes it clear that $\delta(\sqrt{-g} M^{cxy})$ and $\sqg g^{bx}g^{ky}\delta N^{c}_{kb}$ are not equal but become equal on contraction with $g_{xy}$.

Having set up these relations which would allow us to easily hop between the two pairs of HCVs (holographically conjugate variables) that we have, we shall now study the variations of the surface term of the Einstein-Hilbert action on a horizon in terms of our variables.
%#####################################################################

\subsection{Surface Term and its Variation} \label{sec:varLsur}
In this section, we shall calculate the variation of the surface term for an infinitesimal change of the metric. It is a well-known result that the surface term integrated over a horizon in a static metric will give the entropy of the horizon \cite{Padmanabhan:2002sha, Padmanabhan:2004fq, Mukhopadhyay:2006vu, Kolekar:2010dm,Padmanabhan:2012bs,Majhi:2013jpk}. More specifically, it will give $\tau TS$, where $\tau$ is the range of time-integration, $T$ is the Hawking temperature of the horizon and $S$ is the Bekenstein-Hawking entropy of the horizon. Euclidean time arguments are used to replace $\tau$ by $\beta$, the inverse of the horizon temperature, to obtain just $S$. We shall not use this approach. Instead, we shall get rid of the pesky factor of $\tau$ hanging around by defining the surface Hamiltonian for a static metric \cite{Padmanabhan:2012qz,Padmanabhan:2012bs,Majhi:2013jpk} by
\begin{equation}
H_{\textrm{sur}} = -(\partial A_{\textrm{sur}}/\partial\tau) = TS \label{H_sur}
\end{equation}
Then, we need to do our integrations only over the co-ordinates transverse to the horizon. If we then consider a variation of $H_{\textrm{sur}}$, it can be split into two terms: one corresponding to the variation of the metric variable -- like $p\delta q$;  while the other one is like $q\delta p$, coming from the variation of the momentum variable $N^c_{ab}$ or $\sqrt{-g}M^{cab}$. We first give the general expressions for these terms  in terms of metric coefficients and the perturbation  $h_{ab}$. Next, we shall explicitly calculate them on the horizon for a general static spacetime. Interestingly, the first term will lead to $T\delta S$ while the other will give $S\delta T$. Finally, we will generalise this result to an arbitrary null surface.
 
As already described in \ref{prelims}, the surface term in the Einstein-Hilbert Lagrangian is given by $\mLs=\partial_{c}(\sqg V^{c})$, where $V^{c}$ is defined in \ref{Vc}. In order to look at the variations of $H_{\textrm{sur}}$, we need to compute $\df (\sqg V^{c})$.
From \ref{VcM}, we can split the variation of the surface term into two components as
\begin{eqnarray}
\df(\sqg V^{c})&=&-N^{c}_{ab}\df f^{ab}-\df N^{c}_{ab}f^{ab} \label{varVNf} \\
               &=&-\sqg M^{cab}\df g_{ab}-\df(\sqg M_{cab})g_{ab}. 
\label{varVMg}
\end{eqnarray}               
In fact, from \ref{dq_p} and \ref{p_dq}, we know that $N^{r}_{ab}\df f^{ab}=\sqg M^{rab}\df g_{ab}$ and $\df N^{r}_{ab}f^{ab}=\df(\sqg M^{rab})g_{ab}$. We shall work with the variables $f^{ab}$ and $N^{c}_{ab}$ for the moment. In terms of these variables, it is clear that the first term in the above set of equations corresponds to variations of the metric while the second term corresponds to variations of the connection. The variation of the surface Hamiltonian $H_{\textrm{sur}}$ in \ref{H_sur} can then be split into two terms as
\begin{eqnarray}
\df H_{\textrm{sur}} =\frac{1}{16 \pi}\int d^2 x_{\perp} N^{n}_{ab}\df f^{ab} + \frac{1}{16 \pi}\int d^2 x_{\perp} f^{ab} \df N^{n}_{ab} \label{varH_sur}
\end{eqnarray}
where the integration is over the variables transverse to the horizon and the index $n$ refers to the direction normal to the horizon (see \ref{static} for an explicit example). 

To facilitate the calculation, we shall write down expressions, up to linear order, for small changes in the metric. Suppose the change in the metric is of the form $g_{ab}\rightarrow g_{ab} + h_{ab}$ with $h_{ab}$ treated as a first order perturbation. (To be more precise, we should work with $\epsilon h_{ab}$, and retain terms linear in $\epsilon$ and finally set $\epsilon=1$. We shall not bother to do this.)  Under this change, the terms in \ref{varVNf} are evaluated up to linear order in $h_{ab}$ as
\begin{equation}
N^a_{jk}\delta f^{jk} = \sqrt{-g}N^a_{jk}\Big(\frac{1}{2}hg^{jk} - h^{jk}\Big);\label{variation2}
\end{equation}
and
\begin{eqnarray}
\delta N^a_{jk} f^{jk} &=& f^{jk}\Big[-g^{ab}\partial_jh_{bk} + \frac{1}{2} g^{ab}\partial_bh_{jk} + h^{ab}\partial_jg_{bk} - \frac{1}{2}h^{ab}\partial_bg_{jk}\Big] 
\nonumber
\\
&+& \frac{1}{2}f^{ak}\Big[g^{mn}\partial_kh_{mn} - h^{mn}\partial_k g_{mn}\Big]~,
\label{variation3}
\end{eqnarray}
where $h\equiv g^{ab}h_{ab}$.
In the following, we shall evaluate the above terms on the horizon for a general static spacetime.

\subsubsection{A general static spacetime}
\label{static}
 For any static spacetime with a horizon, we can choose an arbitrary $2-$surface and write the line element in the form \cite{Medved:2004ih}: 
\begin{eqnarray}
ds^2 = - N^2 dt^2 + dn^2 + \sigma_{AB}dy^Ady^B~, \label{stmetric}
\label{metric1}
\end{eqnarray}
where the coordinate $n$ corresponds to the spatial direction normal to the specified 2-surface and $\sigma_{AB}$ is the transverse metric on the 2-surface. We shall assume that there exists a Killing horizon determined by the timelike Killing vector $\bf{\xi} = \bf{\partial}_t$ with the location of the horizon  given by the condition $N^2\rightarrow 0$.  We will choose the coordinates such that $n=0$ on the horizon. Then, near the horizon, $N$ and $\sigma_{AB}$ have the expansion \cite{Medved:2004ih},
\begin{equation}
N = \kappa n + {\cal{O}}(n^3);
\,\,\,\
\sigma_{AB} = [\sigma_H(y)]_{AB} + \frac{1}{2}[\sigma_2(y)]_{AB}n^2 + {\cal{O}}(n^3)~.
\label{metric2}
\end{equation}
Here, $\kappa$ is the surface gravity of the horizon. To evaluate \ref{variation2} and \ref{variation3} on the horizon, we shall first calculate them on the $n$=constant surface and then take the limit $n\to0$. The non-zero components of $h_{ab}$ are
$h_{tt} = -2N\delta N; \,\,\ h_{AB} = \delta\sigma_{AB}$
and the relevant non-zero components of $N^a_{bc}$ are
\begin{eqnarray}
N^n_{tt}=-N\partial_n N; \,\,\,\ N^n_{nn} = \frac{\partial_n N}{N} + \frac{1}{2}\sigma^{AB}\partial_n\sigma_{AB}; \,\,\ N^n_{AB} = \frac{1}{2}\partial_n\sigma_{AB}~.
\end{eqnarray}
Using all these in \ref{variation2} and \ref{variation3} we find
\begin{eqnarray}
N^n_{jk}\delta f^{jk} &=& \sqrt{\sigma}\sigma^{AB}\delta\sigma_{AB}\partial_nN - \frac{N}{2}\sqrt{\sigma}\sigma^{AC}\sigma^{BD}\delta\sigma_{CD}\partial_n\sigma_{AB} 
\nonumber
\\
&&+ \frac{N}{2}\sqrt{\sigma}\sigma^{AB}\delta\sigma_{AB}\sigma^{CD}\partial_n\sigma_{CD} + \sqrt{\sigma}\sigma^{AB}\partial_n\sigma_{AB}\delta N~;
\label{variation4}
\end{eqnarray}
and
\begin{equation}
f^{jk}\delta N^n_{jk} = 2\sqrt{\sigma}\partial_n(\delta N) + N\sqrt{\sigma}\sigma^{AB}\partial_n(\delta\sigma_{AB}) - \frac{N}{2}\sqrt{\sigma}\sigma^{AC}\sigma^{BD}\delta\sigma_{CD}\partial_n\sigma_{AB} ~.
\label{variation5}
\end{equation}
Note that the variations that we have considered here are such that the structure of the metric in \ref{metric1} is preserved. We now take the horizon limit and specialize to variations which stay on the horizon, i.e, we take the limits $N\rightarrow 0$, i.e. the horizon limit $n\rightarrow 0$, and $\df N\rightarrow0$. (See the paragraph after \ref{stSdT} for a discussion of the nature of the variations considered.) Using the near-horizon expansion of $N$ from \ref{metric2}, \ref{variation4} and \ref{variation5} become
\begin{equation}
N^n_{jk}\delta f^{jk}|_{H} = 2\kappa\delta(\sqrt{\sigma}); \,\,\,\
f^{jk}\delta N^n_{jk}|_{H} = 2\sqrt{\sigma}\delta\kappa~.
\label{variation7}
\end{equation}
Integrating over the transverse variables and introducing the appropriate numerical factor, we find that the two terms in the variation of the surface Hamiltonian $H_{\textrm{sur}}$ (see \ref{varH_sur}) are given by 
\begin{eqnarray}
\frac{1}{16 \pi}\int d^2 x_{\perp} N^{n}_{ab}\df f^{ab}&=& \frac{\kappa}{2 \pi}\df(\frac{A_\perp}{4})= T \df S;\label{stTdS} \\
\frac{1}{16 \pi}\int d^2 x_{\perp} f^{ab} \df N^{n}_{ab} &=& (\frac{A_\perp}{4})\df(\frac{ \kappa}{2 \pi})= S \df T. 
\label{stSdT}
\end{eqnarray}
In obtaining these results, we have appealed to the zeroth law of black hole thermodynamics \cite{Racz:1995nh,Carter:1973} (See also Eq.(61) in \cite{Medved:2004ih}) and taken $\kappa$ and $\df \kappa$ to be independent of the transverse variables. 

Let us now examine the nature of the variations we have used to obtain our results. As we have already mentioned, the variations are such that they preserve the nature of the metric in \ref{metric1}. So the variations cannot give rise to components of the metric that are zero in \ref{metric1}. Further, the variations that we are considering are differences between quantities evaluated on horizons, which means that we demand $\df (N^2)=0$ or equivalently $\df N=0$. The final horizon need not be at the same location as the initial horizon, as will be the case in the physical example we shall consider in the next section, \ref{sec:fr}.   

Before proceeding further, we address, as an aside, the following question:  We saw in \ref{stSdT} that there is a clear correspondence between the variations $p\df q$ and $q\df p$ on the one hand and $T\df S$ and $S\df T$ on the other hand. How special are the variables $(g_{ab},\sqg M^{cab})$ and $(f^{ab},N^{c}_{ab})$ with respect to this result?
At first sight, it might seem that the separation of the $\df(TS)$ term into $T\df S$ and $S\df T$ terms just corresponds to the separation of the terms with the variation of the metric and the terms with the variation of its first derivatives. We have explicitly verified that this is \textit{not} the case by considering the splitting
\begin{equation}
\df (\sqg V^{c})= -\df (\sqg g^{ab}M^{c}_{ab})=-M^{c}_{ab}\df f^{ab}-f^{ab}\df M^{c}_{ab} \label{VcfM} 
\end{equation}
which did not provide us with the $T\df S+S\df T$ splitting. 
The next natural question would be whether the $\{T\df S,S\df T\}$ structure corresponds to the $\{p\df q,q\df p\}$ structure since $(g_{ab},\sqg M^{cab})$ and $(f^{ab},\sqg N^{c}_{ab})$ are canonically conjugate variables. To answer this question, we looked at another  canonically conjugate pair $(g^{ab},-\sqg M^{c}_{ab})$ (See \ref{Mudd}). The corresponding variation
\begin{equation}
\df (\sqg V^{c})= -\df (g^{ab}\sqg M^{c}_{ab})=-\sqg M^{c}_{ab}\df g^{ab}-g^{ab}\df (\sqg M^{c}_{ab}) \label{VcguuM}
\end{equation}
also \textit{failed} to give us the $\{T\df S,S\df T\}$ splitting, proving that it is not a purely a result of the $\{p\df q,q\df p\}$ structure. 
Thus, $g_{ab}$ and $f^{ab}$ and the corresponding canonical momenta are indeed \textit{special} for our purpose. Although we have not yet discovered a completely satisfactory reason as to why this must be so, we do have some indication that it must be related to the scaling arguments detailed in \ref{app:scaling}. We hope to return to this issue in a future work.
 
We can rewrite \ref{stTdS} in the form of the thermodynamic identity $TdS=dE+dW$.
In order to obtain this identity, we shall borrow some of the results derived in \cite{Kothawala:2009kc}. To begin with, we need to find the variation of $\sqrt{\sigma}$. If $\Lambda$ is the affine parameter corresponding to the tangent vectors of the outgoing null geodesics, then near the horizon we find that
\begin{eqnarray}
\Lambda = \Lambda_H + \frac{1}{2}\kappa n^2 + {\cal{O}}(n^3)~,
\label{affine}
\end{eqnarray}
where $\Lambda=\Lambda_H$ is the location of the horizon. Expressing \ref{metric2} in terms of the affine parameter, we obtain
\begin{eqnarray}
&&N = \sqrt{2\kappa} (\Lambda-\Lambda_H)^{1/2} + {\cal{O}}((\Lambda-\Lambda_H)^{3/4});
\nonumber
\\
&&\sigma_{AB} = [\sigma_H(y,\Lambda_H)]_{AB} + \kappa^{-1}[\sigma_2(y,\Lambda_H)]_{AB}(\Lambda-\Lambda_H) + {\cal{O}}((\Lambda-\Lambda_H)^{3/2})~.
\label{metric3}
\end{eqnarray}
Let us assume that the transverse metric is independent of whatever parameters are present in the metric. (An example is the Schwarzchild metric where the transverse metric is independent of the mass.) Then, the variation of $\sqrt{\sigma}$ would be purely due to a shift in the location of the horizon and can be written as
\begin{eqnarray}
\delta\sqrt{\sigma} \equiv \delta_\Lambda\sqrt{\sigma}= \frac{\partial\sqrt{\sigma}}{\partial\Lambda}\delta\Lambda
= \frac{1}{2}\sqrt{\sigma}\sigma^{AB}\frac{\partial\sigma_{AB}}{\partial\Lambda}\delta\Lambda=\frac{1}{2\kappa}\sqrt{\sigma}\sigma_2 \delta\Lambda~,
\label{sigma}
\end{eqnarray}
where in the last step \ref{metric3} has been used and $\sigma_2\equiv \sigma^{AB}[\sigma_2]_{AB}$. Therefore, we have
\begin{eqnarray}
&&N^n_{jk}\delta f^{jk} = \sqrt{\sigma}\sigma_2 \delta\Lambda~;
\label{variation8}
\end{eqnarray}
Near the horizon, the $nn$ and $tt$ components of the Einstein equations can be written as \cite{Medved:2004ih,Kothawala:2009kc}
\begin{eqnarray}
\frac{1}{2}(\sigma_2 - R_{||}) = 8\pi T^n_n= 8\pi T^t_t~,
\label{Eequation}
\end{eqnarray}
where $R_{||}$ denotes the Ricci scalar on the two-dimensional surfaces of constant $n$ and $t$.
Substituting this in \ref{variation8}, we have
\begin{eqnarray}
N^n_{jk}\delta f^{jk}|_{H} = \sqrt{\sigma}(16\pi T^{*}_{*} + R_{||}) \delta\Lambda~.
\label{variation10}
\end{eqnarray}
Here, $*$ can stand for either $n$ or $t$ because, in our case, $T^n_n=T^{t}_{t}$.
Integrating over the transverse variables and using \ref{stTdS}, we obtain
\begin{equation}
\frac{1}{16 \pi}\int d^2 x_{\perp} N^{n}_{ab}\df f^{ab}=\int d^2 x_{\perp} \sqrt{\sigma}( T^*_* + \frac{R_{||}}{16 \pi})\delta\Lambda  = T \df S.
\end{equation}
The first term in the above equation can be interpreted either as an $\int  PdV $ term or an $\int ( -\rho)dV $ term, where $P$ is the transverse pressure and $\rho$ is the energy density at the horizon, whereas the last term can be interpreted as the variation in energy associated with the horizon $\df E$ (for details, see \cite{Kothawala:2009kc}):
\begin{equation}
\df E = \int d^2 x_{\perp} \sqrt{\sigma}\frac{R_{||}}{16 \pi}\delta\Lambda
\end{equation}
Then, we arrive at either of the following two equations:
\begin{eqnarray}
\df E = T \df S +\int \rho dV ;\\
\df E = T \df S -\int P dV  .  
\end{eqnarray}
Thus, we see that the variations of the surface term on a horizon in a general static spacetime can be given a thermodynamical interpretation. 

We shall now specialize to a spherically symmetric metric in order to gain some more physical insight into our results.

%#######################################################################
\subsubsection{An application: spherically symmetric metric}
\label{sec:fr}
In this subsection, we shall specialize the results of the last section for a spherically symmetric metric of the form
\begin{equation}
ds^2 = -f(r,\lambda)dt^2+ \frac{dr^2}{f(r,\lambda)}+r^2 d\theta^2+r^2 \sin^2\theta d\phi^2 .
\label{fr}
\end{equation}
Here, $\lambda$ is a parameter in the system, like the mass $M$ for a Schwarzchild metric. 
We shall assume that there exists a horizon at $r=r_h$  such that $f(r_h)=0$. 
In order to use the results in the last section, we need to write this metric in the form of \ref{stmetric}. Define a parameter $n$ such that $dn^2=dr^2/f(r)$ and $n=0$ at the horizon. Near the horizon, we can expand $f(r)$ as $f(r)\approx 2 \kappa (r- r_h)$, where $\kappa$ is the surface gravity associated with the horizon at $r_h$. Taking square root and integrating from $r=r_h$ to $r=r$, we obtain $n=\sqrt{2/ \kappa}(r-r_h)^{1/2}$. Comparing with \ref{affine}, we see that $r$ plays the role of the affine parameter here. Having obtained this expression, if we now make the identification $N^{2}=f(r)$ and note that $\sigma_{AB}$ is the metric on the surface of the sphere with radius $r_h$ we have made all the connections necessary to carry over the results in the previous section. But rather than referring back to the results in the last section, we shall rederive these results
(in a slightly different manner)
 since many of the integrals in the last section can be explicitly evaluated in the current case. Also, it will be a good consistency check on our results from the last section. 

For the spherically symmetric metric, the radial component of the surface term in the Einstein-Hilbert action $A_{\textrm{EH}}$ (see \ref{A_EH}, \ref{decomp} and \ref{surVc}), i.e. $\sqg V^{r}/16 \pi$ integrated over $\theta$ and $\phi$ at the 2-surface at $r=r_h$, gives us $-TS$.
\begin{equation}
\frac{1}{16 \pi}\int_{0}^{\pi} d \theta \int_{0}^{2\pi} d \phi \sqg V^{r} =  -TS. \label{Vr-TS}
\end{equation}
From \ref{H_sur}, we can see that this is the negative of the surface Hamiltonian of the horizon.

We shall now look at the variation of the surface Hamiltonian, splitting the variation into a $q\df p$ part and a $p \df q $ part as in \ref{varH_sur}. Our aim is to give a physical example for the results we obtained in \ref{static}. We shall start by considering the variation as being due to the variation of the parameter $\lambda$ due to which the horizon will undergo a shift in position. The condition that the variation is between horizons then means that the variations $\df \lambda$ and $\df r_h$ are connected by the relation $f(r_h,\lambda)=f(r_h+ \df r_h,\lambda+ \df \lambda)=0$, which implies
\begin{equation}
\frac{\partial f}{\partial \lambda}\df \lambda = -f' \df r_h~, \label{horcond}
\end{equation}
a condition that we shall use to simplify our expressions.
The $p \df q$ term in this case is given by
\begin{equation}
\frac{1}{16 \pi}\int_{0}^{\pi} d \theta \int_{0}^{2\pi} d \phi N^{r}_{ab}\df f^{ab}\vbar_{r=r_h}= \frac{r_h}{2}f' \df r_h = (\frac{\kappa}{2 \pi})(\frac{\df(4 \pi r_h^2)}{4})= T \df S, \label{frTdS}
\end{equation}
where we have used the relations that Hawking temperature $T=\kappa/2 \pi=f'/4\pi$ and Bekenstein-Hawking entropy $S= A_\perp /4$.
As can be expected from \ref{Vr-TS}, \ref{varVNf} and \ref{frTdS}, we can obtain the $q \df p$ part as
\begin{equation}
\frac{1}{16 \pi}\int_{0}^{\pi} d \theta \int_{0}^{2\pi} d \phi  f^{ab} \df N^{r}_{ab}\vbar_{r=r_h} = (\frac{4 \pi r_h^2}{4})[(\df \lambda\frac{\partial}{\partial \lambda} + \df r_h\frac{\partial}{\partial r})(\frac{f'}{4 \pi})]\vbar_{r=r_h}  =S \df T. \label{frSdT}
\end{equation}  
\ref{frTdS} can be rewritten making use of the Einstein equations to provide a clearer physical picture. The Einstein equations for the spherically symmetric metric are given by
\begin{align}
G^{t}_{t}=G^{r}_{r}=\frac{rf' + f -1}{r^2}&=8 \pi T^{t}_{t}=8 \pi T^{r}_{r}; \label{E1}\\
G^{\theta}_{\theta}=G^{\phi}_{\phi}=\frac{2f' + rf''}{2r}&= 8 \pi T^{t}_{t}=8 \pi T^{r}_{r}.  \label{E2}
\end{align} 
Substituting in \ref{frTdS} from \ref{E1}, we obtain
\begin{equation}
\frac{1}{16 \pi}\int_{0}^{\pi} d \theta \int_{0}^{2\pi} d \phi N^{r}_{ab}\df f^{ab}\vbar_{r=r_h}= \frac{\df r_h}{2}+ T^{t}_{t}(4 \pi r_h^2 \df r_h)= \frac{\df r_h}{2}+T^{r}_{r}(4 \pi r_h^2 \df r_h)= T \df S. \label{frTdSPdV}
\end{equation}
The factor $4 \pi r_h^2 \df r_h$ is just $dV$, the change in volume when the horizon shifts outward by an amount $\df r_h$. Let us assume that the region outside the horizon contains a perfect fluid at rest. The energy-momentum tensor will be given by
\begin{equation}
T_{ab}=(\rho + P)u_a u_b + P g_{ab}
\end{equation}
and we have $T^{0}_{0}=-\rho$ and $T^{r}_{r}=P$. Then, if $\df r_h/2$ (which is $\df M$ in Schwarzchild case) is interpreted as the change in energy $\df E$, \ref{frTdSPdV} can be written either as
\begin{equation}
\df E = T \df S + \rho \df V  \label{dErhoV}
\end{equation}
or
\begin{equation}
\df E = T \df S - P \df V.     \label{dEPV}
\end{equation}
Among these two interpretations, \ref{dEPV} is in the familiar form of the first law of thermodynamics but \ref{dErhoV} maybe physically more intuitive as it makes clear that there are two contributions to the change in the energy: one contribution from the change in the area of the horizon ($T\df S$ term) and another from the energy density of the matter \textit{engulfed by the horizon} when the horizon expands outward ($\rho dV $ term).

To understand this interpretation clearly, let us consider the special case of the Reissner-N\"{o}rdstrom metric. We can rewrite \ref{frTdSPdV} for this case by substituting for $\df r_h$ in terms of $\df M$ and $\df Q$ using $r_h = r_{\pm}=M\pm \sqrt{M^2 - Q^2}$. We obtain 
\begin{equation}
\frac{1}{16 \pi}\int_{0}^{\pi} d \theta \int_{0}^{2\pi} d \phi N^{r}_{ab}\df f^{ab}\vbar_{r=r_h}=\df M-\frac{Q \df Q}{M\pm \sqrt{M^2 - Q^2}}, 
\end{equation}
where the left-hand side has already been identified with $T\df S$. Further, with $\df E= \df r_h/2$, we can obtain
\begin{equation}
\df E - T \df S =  \frac{Q^2}{8 \pi{r_h}^4}  \left(4 \pi {r_h}^2 \df r_h \right) \label{dErhoVRN1}
\end{equation}
The energy-momentum tensor for an electromagnetic field, which is acting as the source, is given by the expression \cite{Carroll:1997ar}
\begin{equation}
T_{\mu \nu} = \frac{1}{4 \pi} \left( F_{\mu \rho}F^{\phantom{a}\rho}_{\nu}-\frac{1}{4} g_{\mu \nu} F_{\rho \sigma}F^{\rho \sigma} \right)
\end{equation}
For the electromagnetic field of a Reissner-N\"{o}rdstrom metric, there is only one independent non-zero component of the field strength tensor, $F_{tr}=-Q/r^2$.
Evaluating the $T^{0}_{0}$ component at the horizon, we see that \ref{dErhoVRN1} can be rewritten as
\begin{align}
\df E - T \df S = -T^{0}_{0}(4 \pi {r_h}^2 \df r_h)   \label{dErhoVRN}
\end{align}
which leads to $\df E - T \df S = \rho dV$. 
Thus, we reproduce \ref{dErhoV}. Of course, we could also write this relation in the form of \ref{dEPV}. But the current version seems easier to relate to as the right-hand side in \ref{dErhoVRN1} represents the contribution from the electromagnetic energy density that the horizon engulfs as it expands outward. 

%#######################################################################
\subsubsection{Generalization to an Arbitrary Null Surface}
 In the context of emergent gravity paradigm, one extensively uses the concept of a local Rindler frame and local Rindler horizon \cite{Padmanabhan:2009jb}. A local Rindler horizon is essentially a patch of the null surface  in the locally inertial frame. Every local Rindler observer would attribute a temperature and an entropy to the local Rindler horizon. Hence it is natural to extend the above analysis for the metric near an arbitrary null surface. We shall now discuss this formalism.  
   
The metric in the neighbourhood of a null surface is given by 
\begin{eqnarray}
ds^2 =  -2r\alpha du^2 + 2drdu - 2r\beta_Adx^Adu + \mu_{AB}dx^Adx^B~,
\label{null1}
\end{eqnarray}
where $r=0$ corresponds to the null surface. (A detailed construction is presented in \cite{Hollands:2006rj,Morales}). As usual, we will first calculate all the quantities on an $r$-constant surface and then take the $r=0$ limit. The starting point is to find the normal to an $r$-constant surface, which is given by $n_a = N \nabla_a r$ with $N = (2\alpha r+r^2\beta^2)^{-\frac{1}{2}}$. The non-vanishing component of $n_{a}$ is
$n_r = N$.
The surface term on the $r$ = constant surface is given by
\begin{eqnarray}
A_{sur} = \frac{1}{16\pi}\int d^3x\sqrt{h} n_rV^r~,
\label{null3}
\end{eqnarray}
where
$V^r = -(1/g)\partial_b(gg^{rb})$.
For the given metric, it turns out that 
\begin{equation}
\sqrt{h} = \frac{\sqrt{\mu}}{N};\quad
V^r=-\frac{1}{\mu}\Big[\partial_u\mu + \partial_r\{\mu(2r\alpha+r^2\beta^2)\}+\partial_A(\mu r\beta^A)\Big].                                                                                                            \end{equation} 
Therefore,
\begin{eqnarray}
\sqrt{h}n_r V^r &=& - \frac{1}{\sqrt{\mu}}\partial_u\mu - \sqrt{\mu} \Big(2\alpha + 2r\partial_r\alpha+2r\beta^2+2r^2\beta_A\partial_r\beta^A\Big)
\nonumber
\\
&-& \frac{2r\alpha+r^2\beta^2}{\sqrt{\mu}}\partial_r\mu - \sqrt{\mu}r\partial_A\beta^A - \frac{r\beta^A}{\sqrt{\mu}}\partial_A\mu~.
\label{null6}
\end{eqnarray}
Since, the integration variables in \ref{null3} are $u$ and the transverse coordinates $x^A$, it is convenient to take the $r=0$ limit at this stage. This reduces the above equation to the following form:
\begin{eqnarray}
\sqrt{h}n_r V^r = -\frac{1}{\sqrt{\mu}}\partial_u\mu - 2\sqrt{\mu} \alpha~.
\label{null7}
\end{eqnarray}
Assuming the Taylor series expansion:
\begin{eqnarray}
\sqrt{\mu} = \sqrt{\mu^{(0)}(x^A)} + r f(u,x^A)+ {\cal{O}}(r^2)
\label{null8}
\end{eqnarray}
we have $\partial_u\sqrt{\mu} = r \partial_u f(u,x^A) + {\cal{O}}(r^2)$,
and hence
near the null surface we can write:
\begin{equation}
\frac{1}{\sqrt{\mu}}\partial_u\mu = 2\partial_u\sqrt{\mu} = 0; \,\,\,\
\sqrt{\mu} = \sqrt{\mu^{(0)}}~.
\label{null10}
\end{equation}
Therefore \ref{null7} reduces to
$\sqrt{h}n_r V^r =  - 2\sqrt{\mu^{(0)}} \alpha$.
Finally, substituting this in \ref{null3}, we find
\begin{eqnarray}
A_{sur} = - \frac{1}{8\pi}\int du d^2x^A\sqrt{\mu^{(0)}}~ \alpha~.
\label{null12}
\end{eqnarray}
We next expand $\alpha$ in a Taylor series as
\begin{equation}
\alpha (r,u,x^A) = \alpha_0(x^A) + r g(u,x^A)+\dots.
\label{expansionalpha}
\end{equation}
Using this in \ref{null12} and then performing the integration from $u=0$ to $u$ we find
\begin{eqnarray}
A_{sur} = - \frac{\bar{\alpha}_0 A_{\perp}}{8\pi}u~,
\label{sur1}
\end{eqnarray}
where we defined an average surface gravity $\bar{\alpha}_0$ as
\begin{equation}
\bar{\alpha}_0 = \frac{\int d^2x^A \sqrt{\mu^{(0)}}\alpha_0(x^A)}{A_{\perp}}~.
\label{avsurgrav}
\end{equation}
Thus the surface term, calculated on the null surface can be interpreted as $A_{sur} = - u TS$, where $T = \bar{\alpha}_0/2\pi$ and $S=A_{\perp}/4$.

Next we will calculate the terms in \ref{variation3}, which arise due to the  metric variation, on the null surface. We only need to compute $f^{ab}\delta N^r_{ab}$ since the the other term can be identified using \ref{sur1}.
Expanding $f^{jk}\delta N^r_{jk}$ under the metric (\ref{null1}) we obtain,
\begin{eqnarray}
f^{jk}\delta N^r_{jk} &=& 2f^{ur}\delta N^r_{ur}+ f^{rr}\delta N^r_{rr} + 2 f^{rA}\delta N^r_{rA} + f^{AB}\delta N^r_{AB}
\nonumber
\\
&=&2\sqrt{\mu}\delta\Big[-\frac{1}{2}\partial_r{\bar{\alpha}} + \frac{1}{2}\bar{\beta}^A\partial_r\bar{\beta}_A + \frac{1}{4} \mu^{AC}\partial_u\mu_{AC}\Big]
\nonumber
\\
&+& \sqrt{\mu} (2r\alpha+r^2\beta^2)\delta\Big[\frac{1}{2}\mu^{AC}\partial_r\mu_{AC}\Big]
\nonumber
\\
&+& 2r\beta^A\sqrt{\mu} \delta\Big[-\frac{1}{2}\partial_r{\bar{\beta}}_A + \frac{1}{2}\bar{\beta}^C\partial_r\mu_{AC} + \frac{1}{4} \mu^{BD}\partial_A\mu_{BD}\Big]
\nonumber
\\
&+& \sqrt{\mu}\mu^{AB} \delta\Big[ \frac{1}{2}\partial_u\mu_{AB} + \frac{1}{2}(\bar{\beta}^2 - \bar{\alpha})\partial_r\mu_{AB} - D_A\bar{\beta}_B\Big]~,
\label{null14}
\end{eqnarray}
where
\begin{eqnarray}
N^r_{ur} &=& -\Gamma^r_{ur} + \frac{1}{2} \Gamma^a_{au} = \frac{1}{2}\Big(\Gamma^u_{uu} - \Gamma^r_{ur} + \Gamma^A_{Au}\Big)
\nonumber
\\
&=& -\frac{1}{2}\partial_r{\bar{\alpha}} + \frac{1}{2}\bar{\beta}^A\partial_r\bar{\beta}_A + \frac{1}{4} \mu^{AC}\partial_u\mu_{AC}~;
\nonumber
\\
N^r_{rr} &=& \Gamma^a_{ar} = \Gamma^A_{Ar} = \frac{1}{2}\mu^{AC}\partial_r\mu_{AC}~;
\nonumber
\\
N^r_{rA} &=& -\Gamma^r_{rA} + \frac{1}{2} \Gamma^a_{aA} = \frac{1}{2}\Big(\Gamma^u_{uA} - \Gamma^r_{rA} + \Gamma^B_{AB}\Big)
\nonumber
\\
&=& -\frac{1}{2}\partial_r{\bar{\beta}}_A + \frac{1}{2}\bar{\beta}^C\partial_r\mu_{AC} + \frac{1}{4} \mu^{BD}\partial_A\mu_{BD}~;
\nonumber
\\
N^r_{AB} &=& -\Gamma^r_{AB} = \frac{1}{2}\partial_u\mu_{AB} + \frac{1}{2}(\bar{\beta}^2 - \bar{\alpha})\partial_r\mu_{AB} - \frac{1}{2}\Big(D_A\bar{\beta}_B + D_B\bar{\beta}_A\Big)~.
\label{null15}
\end{eqnarray}
with $\bar{\alpha}=-2r\alpha$ and $\bar{\beta}_A = -r\beta_A$ being substituted in the expressions. (To calculate the components of $N^a_{bc}$ we have used \ref{NGamma} and the expressions for the connections evaluated in \cite{Morales}.)
Near the null surface ($r=0 \Rightarrow \delta r=0$), the above result reduces to
\begin{eqnarray}
f^{jk}\delta N^r_{jk} = 2\sqrt{\mu}\delta\Big[-\frac{1}{2}\partial_r{\bar{\alpha}}  + \frac{1}{4} \mu^{AC}\partial_u\mu_{AC}\Big]
+ \sqrt{\mu}\mu^{AB} \delta\Big[\frac{1}{2}\partial_u\mu_{AB}\Big]~.
\label{null16}
\end{eqnarray}
Finally, use of \ref{null8} and \ref{expansionalpha} lead to vanishing of last two terms near $r=0$ and thus we find $f^{jk}\delta N^r_{jk}|_{r=0} = 2\sqrt{\mu^{(0)}}\delta \alpha_0$.
Therefore, the analogue of \ref{stSdT} in this case would be
\begin{eqnarray}
\frac{1}{16\pi}\int d^3x f^{jk} \delta N^r_{jk} = \frac{1}{8\pi}\int d^3x \sqrt{\mu^{(0)}}\delta\alpha_0~,
\label{null17}
\end{eqnarray}
which can be represented as
\begin{equation}
\frac{1}{16\pi }\int d^3x f^{jk} \delta N^r_{jk} = u S\delta T~.
\label{null18}
\end{equation}
Now since $A_{sur} = uTS$, the other term in the variation will be  
\begin{equation}
\frac{1}{16\pi }\int d^3x N^r_{jk}\delta f^{jk} = uT\delta S~.
\label{null19}
\end{equation}
Thus, we have generalized our thermodynamic interpretation to the case of the variation of the surface term in the Einstein-Hilbert action evaluated over an \textit{arbitrary} null surface.

\subsection{Connection with the ADM formalism}

In this section, we shall make some comments relating our formalism with the standard ADM formalism \cite{Arnowitt:1962hi}. We expect to find some  parallels between the two, since the ADM formalism also uses the language of canonical variables. The major difference, of course, is that we have not assumed a particular foliation of the spacetime. In the ADM formalism, we assume a foliation of the spacetime with a family of non-intersecting spacelike hypersurfaces. The dynamics is then generally considered in the volume bounded by two such spacelike surfaces and two timelike hypersurfaces, with one of the timelike hypersurfaces assumed to be at spatial infinity. In the following treatment, we shall borrow the required expressions from Chapter 12 of \cite{gravitation}.

The dynamical variables in the ADM formalism are the components of $h_{\alpha \beta}= g_{\alpha \beta}$, the induced metric on the spacelike surfaces of the foliation. The definition of the term ``canonical momenta'' in the ADM context is different from the definition that we have adopted for our formalism in that the canonical momenta in ADM formalism refers to the the derivatives of the Lagrangian with respect to the \textit{time} derivatives of the metric components and hence correspond to $\sqg M^{0ab}$ in our formalism.  
The ADM Lagrangian, $L_{\textrm{ADM}}$, leads to non-zero canonical momenta conjugate to $h_{\alpha \beta}$ while the momenta corresponding to the other metric components vanish. (This does not happen for $L_{\textrm{quad}}$ and one can find that $M^{000}$ and $M^{00b}$ are in general non-zero.) The canonical momentum corresponding to $h_{\alpha \beta}$ is given by
\begin{equation}
p^{\alpha \beta}= \frac{\partial}{\partial (\partial_{0}{h}_{\alpha \beta})}(\sqg L_{\textrm{ADM}})= -\sqrt{h} (K^{\alpha \beta}-Kh^{\alpha \beta})~. \label{p-h}
\end{equation}
Here, $K^{\alpha \beta}$ represents the contravariant components of the extrinsic curvature $K_{mn}=-h^{a}_{m} \nabla_a n_n$. The extrinsic curvature for this particular foliation is given by
\begin{equation}
K^{mn}= -N h^{am} h^{nb} \Gamma^{0}_{ab} = N h^{am} h^{nb} N^{0}_{ab} \label{Kmn}
\end{equation}
Note that all the indices in the above expression can take only spatial values since $h^{0m}=0$. Taking the trace, we obtain
\begin{equation}
2 \sqh K = -\sqg V^{0} - 2 \sqg n_m n_n M^{0mn} 
\end{equation}
which is the expression relating the Gibbons-Hawking-York counterterm \cite{York:1972sj,Gibbons:1976ue} with the surface term of the Einstein-Hilbert action (see e.g. Exe. 6.3 of \cite{gravitation}). We next write the expression for  $p^{\al \bt}$, given by
\begin{equation}
p^{\al \bt} = 2 \sqg Q^{\al \delta \gamma \beta}_{(h)} \Gamma^{0}_{\gamma \delta} = -2 \sqg Q^{\al \delta \gamma \beta}_{(h)} N^{0}_{\gamma \delta}
\end{equation}
where $2 Q^{a d c b}_{(h)} =  h^{ac} h^{db} - h^{ab} h^{dc}$, in analogy with \ref{lisrq}. 

Having thus made the necessary connections between the variables, we can  look at the variations of the action in the two formalisms. In fact, the variation obtained in Section 12.4.3 of \cite{gravitation} (see Eq. 12.111) is similar in structure to the integrated version of \ref{varRMg} with a $q \df p$ surface term. This variation is given by
\begin{equation}
\int_{\cal{V}}d^4x~ \df( \sqg R) =  \int_{\cal{V}} d^4x \sqrt{-g} G_{ab}\delta g^{ab} +  \int_{\cal{\partial V}} d^3x~ \epsilon h_{ab} \delta p^{ab} 
\end{equation}
where $\epsilon$ is $-1$ on spacelike parts of the boundary $\cal{\partial V}$ and $+1$ on the timelike parts. Here, $h_{ab}$ is the induced metric on the surface of integration and $p^{ab}=- \sqh (K^{ab}-K h^{ab})$, where $K_{ab}$ is the extrinsic curvature corresponding to that surface. \ref{p-h} is a special case in which the surfaces are constant time slices. In obtaining this expression from the variation of the action,  a surface term has been thrown away assuming that the surface of integration is compact.  Comparing with the integrated version of \ref{varRMg}, we obtain
\begin{equation}
-\int_{\cal{\partial V}} d^3x ~g_{ik}\overline{n}_c\delta(\sqg M^{cik}) =  \int_{\cal{\partial V}} d^3x~  ~\epsilon ~h_{ab} \delta p^{ab}~. \label{gdMhdPi}
\end{equation}
Here, $\overline{n}_c$ is the unnormalized normal to the integration surface. For example, if we were integrating over the upper time slice of the boundary of a usual ADM integration volume, an $x^{0}=$constant surface, we would have $\overline{n}_c=-\df^{0}_{c}$, where the minus sign ensures that $\overline{n}^c$ is in the direction of increasing time. The normalized normal in this case would be given by $n_{c}=- N \df^{0}_{c}$. On the other hand, if our integration volume was inside an $r=\textrm{constant}$ surface, we would have $\overline{n}_c=\df^{r}_{c}$ and the normalized normal would be given by $n_{c}=(1/g^{rr}) \df^{0}_{c}$. Thus, we have obtained the correspondence for the $q \df p$ variation term. 

For connecting up with the $p \df q$ variation term, consider the following relation:
\begin{eqnarray}
\int_{\cal{V}}d^4x~ \df( \sqg R) =  \int_{\cal{V}} d^4x \sqrt{-g} G_{ab}\delta g^{ab} + \df\left( \int_{\cal{\partial V}} d^3x~ 2 \sqh K \epsilon \right)-   \int_{\cal{\partial V}} d^3x~\epsilon~  p^{ab} \delta h_{ab}  \label{dRinPidh}
\end{eqnarray}
If the second term was the variation of our usual surface term in the Einstein-Hilbert action, we could have compared the last term directly.with the integral of $N \df f$ term or $M \df g$ term. But we have here the Gibbons-Hawking-York counterterm instead of the surface term in Hilbert action. To obtain the desired relation, we first write down the structure of the usual variation of the Einstein-Hilbert action. This is given by
\begin{equation}
\int_{\cal{V}} d^4x~\df( \sqg R) =  \int_{\cal{V}} d^4x \sqrt{-g} G_{ab}\delta g^{ab} + \df\left( \int_{\cal{\partial V}} d^3x~ \epsilon~ n_i \sqh V^i \right)-  \int_{\cal{\partial V}} d^3x~  n_i \sqh M^{iab} \delta g_{ab} \label{dRusual}
\end{equation}
Comparing \ref{dRusual} with \ref{dRinPidh}, we obtain
\begin{equation}
-  \int_{\cal{\partial V}} d^3x~  \epsilon ~p^{ab} \delta h_{ab} =  -  \int_{\cal{\partial V}} d^3x~  n_i \sqh M^{iab} \delta g_{ab} + \df\left[ \int_{\cal{\partial V}} d^3x~ \epsilon~\sqh(n_i V^i - 2  K)  \right]. \label{genPidhMdg}
\end{equation}
We can now use the following result (see Exe. 6.3 in \cite{gravitation}):
\begin{equation}
V^{a}n_{a}-2 K= 2 h^{ab}\partial_b n_a - n^m h^{ns}\partial_n g_{sm} = - n^m h^{ns}\partial_n g_{sm} \label{VKrel}
\end{equation}
where we have used the result $\partial_b n_{a}=-(\epsilon/2)n_b n_{i}n_{j}~. \partial_a g^{ij}$ and $h^{ab}n_{a}=0$ (see Section 12.4.3 in \cite{gravitation}).
If the metric has no off-diagonal components with respect to the coordinate which labels the surfaces of foliation, the right-hand side of \ref{VKrel} vanishes on the foliation surfaces. In such a case, if we assume that the boundaries of our integration volume other than the foliation surfaces do not contribute, (i.e. in ADM formalism, for example, assuming the integration region is between two time-slices and a time-like surface at spatial infinity where all fields go to zero), we can write \ref{genPidhMdg} as
\begin{equation}
\int_{\cal{\partial V}} d^3x~ \epsilon ~ p^{ab} \delta h_{ab} =    \int_{\cal{\partial V}} d^3x~  n_i \sqh M^{iab} \delta g_{ab}. \label{specPidhMdg}
\end{equation}
This is the desired connection between the $p \df q$ variations; but unlike \ref{gdMhdPi}, this relation is valid only when the
coordinates are chosen such that $2K=n_i V^i$, which can be achieved by demanding a metric block diagonal with respect to the foliation coordinate. (If we take $t=$ constant surfaces for our foliation, for example, then the shift function should vanish.) Further, only the foliation surfaces should contribute to the surface integral. Hence, to summarise, we have a relation between the ``$q\df p$'' variations in ADM formalism and our formalism, \ref{gdMhdPi}:
\begin{equation}
-\int_{\cal{\partial V}} d^3x ~g_{ik}\overline{n}_c\delta(\sqg M^{cik}) =  \int_{\cal{\partial V}} d^3x~  ~\epsilon ~h_{ab} \delta p^{ab}~,
\end{equation}
and, in coordinates in which the metric is block diagonal with respect to the foliation coordinate and non-zero contributions to the surface term come only from the foliation surfaces, we also have the corresponding relation between $p \df q$ variations, \ref{specPidhMdg}:
\begin{equation}
    \int_{\cal{\partial V}} d^3x~  n_i \sqh M^{iab} \delta g_{ab}~ =\int_{\cal{\partial V}} d^3x~\epsilon~  p^{ab} \delta h_{ab} .
\end{equation}
%%%%%%%%%%%%%%%%%%%%%%%%%%%%%%%%%%%%%%%%%%%%%%%%%%%%
\subsection{Action Principle as a Thermodynamical Extremum Principle }

In the previous analysis, we have shown that ${\cal{L}}_{quad}$ can be interpreted as  a Hamiltonian (\ref{sec:Palatini}) -- more precisely a Hamiltonian density-- and the surface integral arising from ${\cal{L}}_{sur}$ leads to $-TS$ on  a horizon  (\ref{sec:varLsur}).  This interpretation leads to the interpretation of the Einstein-Hilbert action as a free energy, $A_{EH} =\tau F= \tau(E-TS)$, 
where $\tau$ is the range of time integration in any static geometry,
which has been explicitly demonstrated for static metrics in  Einstein gravity \cite{Kolekar:2010dm} and also for  Lanczos-Lovelock models \cite{Kolekar:2011bb}. We shall now use this interpretation to formulate the principle of extremisation of gravitational action as a thermodynamic extremum principle.

The variation of the action is given by integrating \ref{varRNf} over a spacetime volume. This variation is given by
\begin{equation}
16 \pi \delta A_{EH} =\int_{\cal{V}} d^4x  \delta(\sqrt{-g}R)= \int_{\cal{V}} d^4x R_{ab}\delta f^{ab}- \int_{\cal{\partial V}} d^3x \sqrt{h} n_i g^{jk}\delta N^i_{jk}, \label{vargr}
\end{equation}
where we have rewritten the volume integral of the total divergence as a surface integral.

Now consider the variation of the action on-shell. For pure gravity ($R_{ab}=0$), the above variation reduces to
\begin{equation}
 \delta A_{EH} =  - \frac{1}{16 \pi}\int_{\cal{\partial V}} d^3x \sqrt{h} n_i g^{jk}\delta N^i_{jk} =  -\frac{1}{16 \pi}\int_{\cal{\partial V}} d^3x  f^{jk}\delta N^X_{jk}~,
\label{EHO1}
\end{equation}
on the $X=$constant surface. Here $X=n$ for a static spacetime while $X=r$ for null metric. From \ref{sec:varLsur}, we know that the right-hand side, when evaluated on a horizon, can be interpreted as a $-\tau S\delta T$ term. Therefore, using $A_{EH} = \tau(E-TS)$ as has been argued, we find that the \ref{EHO1} can be written as a thermodynamic identity:
\begin{equation}
\delta (E-TS) = -S\delta T; \quad \mathrm{i.e.}, \ \delta E = T \delta S \label{varA-therm}
\end{equation}

Next we shall consider the inclusion of the matter action $A_m$. The usual matter Lagrangians are independent of the derivatives of the metric and hence the variation with respect to the metric will not involve any surface terms. The variation of the matter Lagrangian on varying the metric is then written in the form
\begin{equation}
\delta A_{m} = \frac{1}{2} \int_{\cal{V}} d^4x \sqg T^{ab} \df g_{ab}  = -\frac{1}{2} \int_{\cal{V}} d^4x \sqg \overline{T}_{ab} \df f^{ab}. \label{varM}
\end{equation}
Here, $\overline{T}_{ab}= T_{ab}-(g_{ab}/2)T^{i}_{i}$ and $T_{ab}$ is the energy momentum tensor corresponding to the matter field under consideration. 
Then, from \ref{vargr} and \ref{varM}, we can impose the on-shell condition $R_{ab} = 8\pi G \overline{T}_{ab}$ and obtain
\begin{equation}
\delta [A_{EH} + A_m] =   -\int_{\cal{\partial V}} d^3x  f^{jk}\delta N^X_{jk}~.
\label{EHO2}
\end{equation}
If the matter is perfect fluid, then the matter action, with the on-shell condition, can be expressed as
\begin{equation}
A_m = \int \sqrt{-g}d^4x P                          
\end{equation} 
where $P$ is the pressure of the fluid \cite{PhysRevD.2.2762,Brown:1992kc,PhysRevD.86.087502}. For the case of a static spacetime with $P$ independent of $f^{ab}$, the time integration can be performed to give a factor $\tau$ and the variation of the matter action will reduce to $\tau P\delta V$, where $V$ stands for the three-dimensional volume of a time-constant slice. Hence, the left hand side of \ref{EHO2} can be given a thermodynamic interpretation as $\tau \delta(E-TS)+\tau P\delta V$. Thus, \ref{varA-therm}, in a static space time with the inclusion of matter fields in the form of a perfect fluid with $P=$constant, becomes
$\delta E = T\delta S - P\delta V,$
which is the thermodynamic identity on the horizon obtained earlier. We thus see that the variation of the Einstein-Hilbert action allows for a straightforward thermodynamic interpretation.

There is, however, a nicer way of interpreting the gravitational action principle in thermodynamic language, along the following lines. We first note that, if --- instead of demanding $\delta [A_{EH} + A_m]=0$ --- we demand the condition in \ref{EHO2} we will get the field equations. In such a formulation, we can use any spacetime region $\mathcal{V}$ and its boundary $\partial\mathcal{V}$. Consider now a spacetime region bounded by null surfaces. Then the surface term on the right hand side of 
\ref{EHO2} can be interpreted as giving $S\delta T$ based on our earlier result in \ref{null18}. Hence, the condition that the surface term vanishes is equivalent to the condition that the variations keep the temperature of the null surfaces, as perceived by the local Rindler observers, constant during the variation. This gives a very direct thermodynamic interpretation of the gravitational action principle provided we formulate it in a region bounded by null surfaces.
We hope to explore this in detail in a future publication.
%%%%%%%%%%%%%%%%%%%%%%%%%%%%%%%%%%%%%%%%%%%%%%%

\section{Conclusions}
\label{concl}
The variational principle in general relativity is somewhat peculiar. The key reason is the presence of second derivatives of the metric in the Einstein-Hilbert action, which causes the surface term in the variation to contain variations of the metric and its derivative. Thus, in order to get the Einstein's equation by the usual variational method, we have to fix the metric as well as its normal derivative. The main problems with fixing \textit{both} the metric and the normal derivatives are:
(a) The values at which we fix the metric and its derivative at the boundaries might not turn out to be compatible with the equations of motion
derived.
(b) If we extend our theory to the quantum domain, we should refrain from fixing both the metric and its normal derivative on a spacelike hypersurface to avoid conflict with the uncertainty principle. 
One common method for dealing with this issue is to throw the second derivatives into a surface term and then work with the remaining (coordinate dependent) quadratic Lagrangian $\mLq$ (see \ref{decomp}), in which case one will have to fix just the variation of the metric on the surface.  The other common procedure is to use a counterterm to cancel the variation of the surface term \cite{Gibbons:1976ue}. All these procedures, one must admit,  appear rather contrived. 

In this paper, we discuss an alternate prescription. We may be in better shape conceptually by interpreting the generally covariant Lagrangian (viz. the Einstein-Hilbert Lagrangian) as a momentum space Lagrangian
and fixing the canonical momenta on the boundary. We can do this in terms of the metric $g_{ab}$ and its canonical momenta with respect to the quadratic Lagrangian, $\sqg M^{cab}$, defined in \ref{M3}. This suggests that general relativity is better represented as a theory in the space of the canonical momenta $M^{cab}$. The immediate direction suggested by this realization is to  apply the technique of momentum-space path integrals to general relativity, a direction which we intend to explore in the future.

In the process, we discovered the surprising fact that this approach works only with $g_{ab}$ but not with $g^{ab}$ and its corresponding canonical momenta. In the case of $g^{ab}$, the surface variation is found to contain the variation of $g^{ab}$ \textit{as well as} its canonical momenta. We show how this is related to the $-\partial(pq)$ structure of the surface term in the case of the variable $g_{ab}$ and proceed to show how this structure can be obtained by simple scaling arguments applicable for homogeneous functions. These arguments also allow us to discover another variable  $f^{ab}=\sqg g^{ab}$, with $N^{c}_{ab}$ representing the corresponding canonical momenta. The use of $N^{c}_{ab}$ makes it easy to see that we are actually fixing the variation of the connection on the boundary, a fact which needs nontrivial calculation to discover when working with $g_{ab}$ and $M^{cab}$. Further, as already noted in the literature before (see \cite{Eddington:1924,Schrod:1950,Einstein:1955ez,Kijowski:97}), many formulas simplify if we work with $f^{ab}$ and $N^{c}_{ab}$. 

The most surprising result of our investigation was the connection between these holographically conjugate variables (HCVs) and thermodynamic quantities pertaining to the null surfaces which act as local Rindler horizons. The  surface term in the Einstein-Hilbert action, when integrated over a horizon, gives us the Bekenstein-Hawking entropy of the horizon when the range of time integration is fixed by periodicity in Euclidean sector. Without using the Euclidean time integration, and in fact removing the time integration altogether, the integral of the surface term over the space variables on the horizon gives us \cite{Padmanabhan:2012qz,Padmanabhan:2012bs,Majhi:2013jpk} the heat content $H_{\rm sur} =  TS$. In a variation of this integral, it was seen that the term $T\df S$ is the term with the variation of $g_{ab}$ (or $f^{ab}$) and the term with $S\df T$ is the term with the variation of the corresponding canonical momentum. This result holds: (a)near a horizon in an arbitrary (i.e., not necessarily spherically symmetric) static spacetime and (b) near any null surface which acts as a local Rindler horizon.

We believe that this is a very strong result. We know that the variations of $T$ and $S$ are related to the variations of the surface gravity $\kappa$ and the area of the horizon respectively. Since $\kappa$ is related to the derivative of the $g_{00}$ component of the metric along the direction normal to the horizon, it is clear that the variation of the temperature cannot be given by the term with the variation of $g_{ab}$ or $f^{ab}$ and must be contained in the term with the variation of the canonical momenta. But it is not clear why this canonical momenta term does not contribute to the variation of the entropy. To drive this point home, we carried out variations with certain other sets of variables (see \ref{VcfM} and \ref{VcguuM} and the accompanying discussion) and observed that we do \textit{not} obtain this separation between the entropy and the temperature. Although we do not yet have a clear line of reasoning to offer as to why this must be so, we think that the explanation might be related to the scaling properties that were used to  arrive at the variable $f^{ab}$. Further, we were able to obtain this result without using the Euclidean time method. This seems to suggest that the method of reducing the integral of surface term on a horizon to entropy using Euclidean time arguments might not be necessary.

There are many directions to proceed forward from the work in this paper. It is not clear to us why the special variables that we discovered through scaling turn out to be the ones that are related to thermodynamic variables. This connection should be further explored. As already mentioned, another direction of work would be to try to develop a momentum space path integral approach to general relativity. Extension of our results to the case of stationary, or more ambitiously, time-dependent metrics and \LL\ models  is another obvious line of attack. The simplicity of the scaling argument that led us to discover the special nature of the variables $g_{ab}$ and $f^{ab}$, the naturalness of our prescription for the variational approach to general relativity compared to prescriptions existing in the literature as well as the intriguing connection with thermodynamic variables on a horizon which is highly unlikely to be accidental, suggests that these directions of research would prove to be quite fruitful if pursued. 
 
%%%%%%%%%%%%%%%%%%% ACKNOWLEDGMENTS %%%%%%%%%%%%%%%%%%%

\section*{Acknowledgments}

The research of TP is partially supported by J.C.Bose research grant of DST, India. KP is supported by the Shyama Prasad Mukherjee Fellowship from the Council of Scientific and Industrial Research (CSIR), India. KP would like to thank Jose Mathew for discussions on classical mechanics and Sanved Kolekar for his comments.

%%%%%%%%%%%%%%%%%%% APPENDICES %%%%%%%%%%%%%%%%%%%
\appendix

\labelformat{section}{Appendix #1}

%#######################################################################
\section{Finding Holographically Conjugate Variables through Scaling Arguments}
\label{app:scaling}
\renewcommand{\theequation}{A.\arabic{equation}}
\setcounter{equation}{0}
In this section, we shall try to find alternatives to $g_{ab}$ for which a relation of the form in \ref{holgdd} holds. More explicitly, for \ref{holgen} (which we reproduce below):
\begin{equation}
q_A F^A=(\lambda + \mu)L-\partial_i \left[q_A \frac{\partial L}{\partial(\partial_i q_A)}\right], \label{holgen1}
\end{equation}
we should have $q_A F^A = \sqg R$ and $\lambda+\mu=1$.
We shall restrict ourselves to variables that can be obtained by the so-called point transformations \cite{Goldstein} i.e transformations where the new variables depend on the old variables, but not on their derivatives i.e
\begin{equation}
h = h(g_{ab},x_i);\phantom{parattu} \frac{\partial h}{\partial( \partial_c g_{ab})} =0. 
\end{equation}
Further, we shall assume that the transformation has no explicit dependence on spacetime coordinates i.e $h(g_{ab},x_i)=h(g_{ab})$.
Here we have not specified the index structure of the variable $h$, but it has to be a 2-indexed symmetric object in order to hold the degrees of freedom initially present in the metric. 
First, let us look at the left-hand side of \ref{holgen1}. For a point transformation from the variable $q$ to the variable $s$, we have the result that 
\begin{equation}
\frac{d}{dt}\left(\frac{\partial L}{\partial \dot{s}_j}\right)-\frac{\partial L}{\partial s_j}=\left[\frac{d}{dt}\left(\frac{\partial L}{\partial \dot{q}_i}\right)-\frac{\partial L}{\partial s_j}\right] \frac{\partial q_i}{\partial s_j}.
\end{equation}
Generalizing to our case, we have
\begin{equation}
\partial_c\left(\frac{\partial L}{\partial (\partial_c g_{ab})}\right)-\frac{\partial L}{\partial g_{ab}}=\left[\partial_c\left(\frac{\partial L}{\partial (\partial_c h)}\right)-\frac{\partial L}{\partial h}\right] \frac{\partial h}{\partial g_{ab}}. \label{pointEuler}
\end{equation}
We shall now assume that $h$ is a homogeneous function of the components of $g_{ab}$ i.e if $g_{ab}\rightarrow \alpha g_{ab}$, then $h\rightarrow \alpha^{k}h$ for some constant $k$. If we now contract both sides of \ref{pointEuler} with $g_{ab}$ and use Euler's theorem for homogeneous functions, we obtain
\begin{equation}
g_{ab}\left[\partial_c\left(\frac{\partial L}{\partial (\partial_c g_{ab})}\right)-\frac{\partial L}{\partial g_{ab}}\right]=k h\left[\partial_c\left(\frac{\partial L}{\partial (\partial_c h)}\right)-\frac{\partial L}{\partial h}\right]. 
\end{equation}
In terms of the notation in \ref{holgen1}, we can state this result as follows. If we transform from a set of variables $q_A$ to another set $f_A$, such that the $f_A$ do not dependent on the derivatives of $q_A$ and are homogeneous of degree $k$ in $q_A$, then
\begin{equation}
q_A F_q^{A}\longrightarrow k f_A F_{f}^{A}, \label{qFfF} 
\end{equation}
where we have used the subscript in the Euler-Lagrange function to denote the variable that has been used.

Next we shall look at the right-hand side of \ref{holgen1}. Under a point transformations, it is easy to see that the value of $\lambda$, the degree of derivatives of the variables, remains a constant. A fallacious argument for finding the change in $\mu$ is the following. $f_A$ being homogeneous in $q_A$ of degree $k$ and $L$ being homogeneous in $q_A$ of degree $\mu$, a change $q_A\rightarrow \alpha q_A$ would correspond to a change $f_A\rightarrow \alpha^k f_A$ and a change $L\rightarrow \alpha^\mu L$. Therefore, under a change $f_A\rightarrow \alpha f_A$, we should have $q_A\rightarrow \alpha^{1/k} q_A$ and $L\rightarrow \alpha^{\mu/k} L$ and we can conclude that the degree of $L$ in the variable $f_A$ is $\mu/k$. But as is clear from \ref{Holoscaling}, this argument fails for the transformation $g_{ab}\rightarrow g^{ab}$. To find out the error in the above argument and to derive the correct result, consider a general term in the Lagrangian of the form $(q_A)^{\mu}(\partial_i q_B)^{\lambda}$. The notation $(q_A)^\mu$ here corresponds to a term of the form $\{(q_A q_B ....)\rightarrow\mu \textrm{ terms}\}$ and $(\partial_i q_B)^{\lambda}$ corresponds to a term of the form $\{(\partial_i q_A \partial_j q_B ....)\rightarrow\lambda \textrm{ terms}\}$. In terms of the new variables $f_A$, this term becomes 
\begin{equation}
(q_A)^{\mu}(\partial_i q_B)^{\lambda}=(q_{A}[f_C])^{\mu}(\frac {\partial q_B}{\partial f_C} \partial_i f_C)^{\lambda},
\end{equation}
where we have made use of the fact that $q_A$ does not depend on the derivatives of $f_B$. Now, this assumption also tells us that $(\partial q_B / \partial f_C)$ is a function of $f_A$ alone and not its derivatives. Thus, the factor containing derivatives of $f_A$ is ($\partial f_C)^{\lambda}$ confirming that the degree in derivatives does not change under the transformation. The factor that depends only on the variables $f_A$ and not on its derivatives is
\begin{equation}
(q_{A}[f_C])^{\mu}(\frac {\partial q_B}{\partial f_C})^{\lambda}.
\end{equation}
We have considered $f_A$ to be a homogeneous function of $q_B$, and $q_B$ only, of degree $k$. Since we are conserving the degrees of freedom, we should be able to invert these functions and express $q_B$ in terms of $f_A$ as homogeneous functions of degree $1/k$ in $f_A$. Then, $(\partial q_B)/(\partial f_C)$ will be a homogeneous function of $f_A$ of degree $(1/k)-1$. Hence, the above function will be a homogeneous function of $f_A$ alone, and of degree $(1/k)\mu+[(1/k)-1]\lambda$.

It is straightforward to generalize our results for $(q_A)^{\mu}(\partial q_B)^{\lambda}$ to $L$ and conclude that, under a transformation from variables $q_A$ to variables $f_B$ homogeneous of degree $k$ in $q_A$ and independent of the derivatives of $q_A$, we shall have 
\begin{equation}
\mu\rightarrow \frac{\mu}{k}+\left(\frac{1}{k}-1\right)\lambda;\phantom{p} \lambda\rightarrow \lambda \phantom{aa};\quad \mathrm{i.e.}, \phantom{aa} \lambda + \mu \rightarrow \frac{\lambda + \mu}{k}. \label{lmtransf}
\end{equation}
Thus, from \ref{holgen1}, \ref{qFfF} and \ref{lmtransf}, the form of \ref{holgdd} is conserved \textit{only if} we transform to a variable which is homogeneous of degree $k=1$ in $g_{ab}$. This obviously breaks down for $g^{ab}$, which has $k=-1$. The determinant of $g_{ab}$, $g$, does not have the required number of degrees of freedom and has $k=4$ in four dimensions and is also unsuitable. But the variable 
\begin{equation}
f^{ab}=\sqg g^{ab}   \nonumber
\end{equation}
has the required number of degrees of freedom and has $k=1$, thus providing a useful alternative, contravariant in its two indices,  to $g_{ab}$. 
%#######################################################################
\section{Useful properties and relations pertaining to $f^{ab}$}
\label{app:fab}
\renewcommand{\theequation}{B.\arabic{equation}}
\setcounter{equation}{0}
In this appendix, we shall list out some useful properties and relations pertaining to $f^{ab}=\sqg g^{ab}$. Its determinant is given by
\begin{equation}
f=\textrm{det}(f^{ab})=\textrm{det}(g_{ab})=g.
\end{equation}
Hence, all the $\sqg$ factors that hang around in expressions can be replaced by $\sqrt{-f}$. In order to use the well-known formulas used in variation of the gravitational action while working with $f^{ab}$, we need to relate the variation of $g^{ab}$ to the variation of $f^{ab}$. This relation is given by  
\begin{eqnarray}
\df g^{lm}=\frac{\df f^{lm}}{\sqrt{-f}}-\frac{f^{lm}f_{ab}\df f^{ab}}{2(\sqrt{-f})^{3}}=\frac{B^{lm}_{ab}\df f^{ab}}{\sqrt{-f}}, \label{guuinf}
\end{eqnarray}
where $B^{lm}_{ab}\equiv (1/2)(\df^{l}_{a}\df^{m}_{b}+\df^{l}_{b}\df^{m}_{a})-(1/2)g^{lm}g_{ab}=\df^{l}_{(a}\df^{m}_{b)}-(1/2)g^{lm}g_{ab}$. We shall take $B^{lm}_{ab}$ to be $B^{lm}_{\phantom{lm}ab}$ so that $B_{lmab}= g_{l(a}g_{mb)}-(1/2)g_{lm}g_{ab}$. $B^{lm}_{ab}$ satisfies the relation
\begin{equation}
B^{lm}_{ab}B^{ab}_{cd}=\df^{l}_{(c}\df^{m}_{d)}.
\label{Bnorm}
\end{equation}
This relation is valid even if we remove the explicit symmetrisation in both $B^{lm}_{bk}$ and the right-hand side of this relation. Using this relation, we can easily invert \ref{guuinf} and obtain (see Chapter 6 in \cite{gravitation}) 
\begin{equation}
\df f^{lm}=\sqg B^{lm}_{ab}\df g^{ab}. \label{dfwithdg}
\end{equation}
Hence, we have, for any two-indexed object $X_{lm}$
\begin{eqnarray}
X_{lm}\df g^{lm}&=&\frac{1}{\sqg}[X_{ab}-\frac{1}{2}g_{ab}g^{lm}X_{lm}] \df f^{ab}\nonumber\\
&=&\frac{1}{\sqg}[X_{ab}-\frac{1}{2}g_{ab}X] \df f^{ab}
\end{eqnarray}
Therefore, 
\begin{equation}
\sqg G_{ab} \df g^{ab} = R_{ab} \df f^{ab}. \label{GgRf} 
\end{equation}
The Euler-Lagrange function (see \ref{Euler-deriv}) $F_{ab}$ for the variable $f^{ab}$ is
\begin{eqnarray}
F_{ab}\equiv \frac{\partial(\sqg L_{\textrm{quad}} )}{\partial f^{ab}}-\partial_{c}[\frac{\partial (\sqg L_{\textrm{quad}} )}{\partial( \partial_c f^{ab})}]
= R_{ab}.
\end{eqnarray}
This expression can be arrived at either by explicit computation or by staring at \ref{varRNf}.

%#######################################################################
\section{Proof of the Conservation Equation $\partial_{i}(t^{i}_{k}-16 \pi\sqg T^{i}_{k})=0$}
%%%%%%%%%%%%%%%%%%%%%%%%%%%%%%%%%%%%%%%%%%%%%%%%
\label{app:conserv}
\renewcommand{\theequation}{C.\arabic{equation}}
\setcounter{equation}{0}

Consider the object $\partial_i t^{i}_{k}$. We have,using the definition in \ref{tikf},
\begin{eqnarray}
\partial_i t^{i}_{k}&=&\partial_i\left[\frac{\partial(\sqg L_{quad})}{\partial(\partial_i f^{ab})}\partial_{k}f^{ab}- \df^{i}_{k} \sqrt{-f} L_{quad}\right] \nonumber\\
&=& \partial_i \left[\frac{\partial(\sqg L_{quad})}{\partial(\partial_i f^{ab})}\partial_{k}f^{ab}\right]- \frac{\partial(\sqg L_{quad})}{\partial f^{ab}}\partial_k f^{ab}- \frac{\partial(\sqg L_{quad})}{\partial(\partial_i f^{ab})}\partial_k \partial_i f^{ab} \nonumber  \\
&=& \left\{\partial_i \left[\frac{\partial(\sqg L_{quad})}{\partial(\partial_i f^{ab})}\right]- \frac{\partial(\sqg L_{quad})}{\partial f^{ab}}\right\}\partial_k f^{ab} \nonumber \\
&=& -R_{ab} \partial_k f^{ab}
\end{eqnarray}
It is clear from the above equation that, in the absence of matter, the object $t^{i}_{k}$ will be conserved  once we impose the equations of motion.

To generalize to the case with matter present, start from the Bianchi identity $\sqg \nabla_{i} G^{i}_{k}=0$. Then, we have $\sqg \nabla_{i} R^{i}_{k} = (\sqg/2) \nabla_k R$, implying
\begin{eqnarray}
2 \partial_i (\sqg R^{i}_{k}) = \sqg g^{ab} \nabla_k R_{ab} + \sqg R^{ab}\partial_k g_{ab} =  f^{ab} \partial_k R_{ab} ~.
\end{eqnarray}
Hence, we have
\begin{align}
\partial_i t^{i}_{k} &= -R_{ab} \partial_k f^{ab} 
= -\partial_k (\sqg R) + f^{ab} \partial_k R_{ab} \nn \\
&=  -\partial_k (\sqg R) + 2 \partial_i (\sqg R^{i}_{k}) 
= 2 \partial_{i}\left[\sqg (R^{i}_{k}-\frac{\df^{i}_{k}}{2}R)\right] \nn \\
&= 2 \partial_i (\sqg G^{i}_{k} ) 
= 16 \pi~ \partial_i (\sqg T^{i}_{k} ) 
\end{align}
where we have used the Einstein equations $G^{i}_{k}=8 \pi  T^{i}_{k}$ in the last step. Thus, we obtain the general conservation equation
\begin{equation}
\partial_i (t^{i}_{k} - 16 \pi  \sqg T^{i}_{k}) = 0 
\end{equation}

%%%%%%%%%%%%%%%%%%% BIBLIOGRAPHY %%%%%%%%%%%%%%%%%%%
% \clearpage
% \newpage
\bibliography{mybibliography-gravity}

\providecommand{\href}[2]{#2}\begingroup\raggedright\begin{thebibliography}{10}

\bibitem{Bekenstein:1972tm}
J.~Bekenstein, ``{Black holes and the second law},'' {\em Lett. Nuovo Cimento
  Soc. Ital. Fis.} {\bf 4} (1972)
737--740.
%%CITATION = NCLTA,4,737;%%.

\bibitem{Bekenstein:1973ur}
J.~D. Bekenstein, ``{Black holes and entropy},'' {\em Phys.Rev.} {\bf D7}
  (1973)
2333--2346.
%%CITATION = PHRVA,D7,2333;%%.

\bibitem{Bekenstein:1974ax}
J.~D. Bekenstein, ``{Generalized second law of thermodynamics in black hole
  physics},'' {\em Phys.Rev.} {\bf D9} (1974)
3292--3300.
%%CITATION = PHRVA,D9,3292;%%.

\bibitem{Hawking:1974sw}
S.~Hawking, ``{Particle Creation by Black Holes},'' {\em Commun.Math.Phys.}
  {\bf 43} (1975)
199--220.
%%CITATION = CMPHA,43,199;%%.

\bibitem{Hawking:1976de}
S.~Hawking, ``{Black Holes and Thermodynamics},'' {\em Phys.Rev.} {\bf D13}
  (1976)
191--197.
%%CITATION = PHRVA,D13,191;%%.

\bibitem{Wald:1999vt}
R.~M. Wald, ``{The thermodynamics of black holes},'' {\em Living Rev.Rel.} {\bf
  4} (2001) 6,
\href{http://www.arXiv.org/abs/gr-qc/9912119}{{\tt gr-qc/9912119}}.
%%CITATION = GR-QC/9912119;%%.

\bibitem{Padmanabhan:2003gd}
T.~Padmanabhan, ``{Gravity and the thermodynamics of horizons},'' {\em
  Phys.Rept.} {\bf 406} (2005) 49--125,
\href{http://www.arXiv.org/abs/gr-qc/0311036}{{\tt gr-qc/0311036}}.
%%CITATION = GR-QC/0311036;%%.

\bibitem{Padmanabhan:2009vy}
T.~Padmanabhan, ``{Thermodynamical Aspects of Gravity: New insights},'' {\em
  Rept. Prog. Phys.} {\bf 73} (2010) 046901,
  \href{http://www.arXiv.org/abs/0911.5004}{{\tt 0911.5004}}.

\bibitem{Padmanabhan:2012qz}
T.~Padmanabhan, ``{Structural Aspects Of Gravitational Dynamics And The
  Emergent Perspective Of Gravity},'' {\em AIP Conf.Proc.} {\bf 1483} (2012)
  212--238,
\href{http://www.arXiv.org/abs/1208.1375}{{\tt 1208.1375}}.
%%CITATION = ARXIV:1208.1375;%%.

\bibitem{Sakarov}
A.~D. Sakharov, ``{},'' {\em Sov. Phys. Dokl.} {\bf 12} (1968) 1040.

\bibitem{Hu:1996gk}
B.~Hu, ``{General relativity as geometrohydrodynamics},''
\href{http://www.arXiv.org/abs/gr-qc/9607070}{{\tt gr-qc/9607070}}.
%%CITATION = GR-QC/9607070;%%.

\bibitem{Jacobson:1995ab}
T.~Jacobson, ``{Thermodynamics of space-time: The Einstein equation of
  state},'' {\em Phys.Rev.Lett.} {\bf 75} (1995) 1260--1263,
\href{http://www.arXiv.org/abs/gr-qc/9504004}{{\tt gr-qc/9504004}}.
%%CITATION = GR-QC/9504004;%%.

\bibitem{Barcelo:2005fc}
C.~Barcelo, S.~Liberati, and M.~Visser, ``{Analogue gravity},'' {\em Living
  Rev.Rel.} {\bf 8} (2005) 12,
\href{http://www.arXiv.org/abs/gr-qc/0505065}{{\tt gr-qc/0505065}}.
%%CITATION = GR-QC/0505065;%%.

\bibitem{Volovik}
G.~E. Volovik, ``{The universe in a helium droplet},'' {\em Oxford University
  Press} (2003).

\bibitem{Padmanabhan:2002sha}
T.~Padmanabhan, ``{Classical and quantum thermodynamics of horizons in
  spherically symmetric space-times},'' {\em Class.Quant.Grav.} {\bf 19} (2002)
  5387--5408,
\href{http://www.arXiv.org/abs/gr-qc/0204019}{{\tt gr-qc/0204019}}.
%%CITATION = GR-QC/0204019;%%.

\bibitem{Kothawala:2009kc}
D.~Kothawala and T.~Padmanabhan, ``{Thermodynamic structure of Lanczos-Lovelock
  field equations from near-horizon symmetries},'' {\em Phys. Rev.} {\bf D79}
  (2009) 104020, \href{http://www.arXiv.org/abs/0904.0215}{{\tt 0904.0215}}.

\bibitem{Padmanabhan:2009jb}
T.~Padmanabhan, ``{A Physical Interpretation of Gravitational Field
  Equations},'' {\em AIP Conf.Proc.} {\bf 1241} (2010) 93--108,
\href{http://www.arXiv.org/abs/0911.1403}{{\tt 0911.1403}}.
%%CITATION = ARXIV:0911.1403;%%.

\bibitem{Padmanabhan:2007en}
T.~Padmanabhan and A.~Paranjape, ``{Entropy of null surfaces and dynamics of
  spacetime},'' {\em Phys.Rev.} {\bf D75} (2007) 064004,
\href{http://www.arXiv.org/abs/gr-qc/0701003}{{\tt gr-qc/0701003}}.
%%CITATION = GR-QC/0701003;%%.

\bibitem{Padmanabhan:2007xy}
T.~Padmanabhan, ``{Dark energy and gravity},'' {\em Gen.Rel.Grav.} {\bf 40}
  (2008) 529--564,
\href{http://www.arXiv.org/abs/0705.2533}{{\tt 0705.2533}}.
%%CITATION = ARXIV:0705.2533;%%.

\bibitem{Padmanabhan:2009kr}
T.~Padmanabhan, ``{Equipartition of energy in the horizon degrees of freedom
  and the emergence of gravity},'' {\em Mod.Phys.Lett.} {\bf A25} (2010)
  1129--1136,
\href{http://www.arXiv.org/abs/0912.3165}{{\tt 0912.3165}}.
%%CITATION = ARXIV:0912.3165;%%.

\bibitem{Padmanabhan:2010xh}
T.~Padmanabhan, ``{Surface Density of Spacetime Degrees of Freedom from
  Equipartition Law in theories of Gravity},'' {\em Phys.Rev.} {\bf D81} (2010)
  124040,
\href{http://www.arXiv.org/abs/1003.5665}{{\tt 1003.5665}}.
%%CITATION = ARXIV:1003.5665;%%.

\bibitem{Padmanabhan:2004fq}
T.~Padmanabhan, ``{Holographic gravity and the surface term in the
  Einstein-Hilbert action},'' {\em Braz.J.Phys.} {\bf 35} (2005) 362--372,
\href{http://www.arXiv.org/abs/gr-qc/0412068}{{\tt gr-qc/0412068}}.
%%CITATION = GR-QC/0412068;%%.

\bibitem{Mukhopadhyay:2006vu}
A.~Mukhopadhyay and T.~Padmanabhan, ``{Holography of gravitational action
  functionals},'' {\em Phys. Rev.} {\bf D74} (2006) 124023,
  \href{http://www.arXiv.org/abs/hep-th/0608120}{{\tt hep-th/0608120}}.

\bibitem{Kolekar:2010dm}
S.~Kolekar and T.~Padmanabhan, ``{Holography in Action},'' {\em Phys. Rev.}
  {\bf D82} (2010) 024036, \href{http://www.arXiv.org/abs/1005.0619}{{\tt
  1005.0619}}.

\bibitem{Padmanabhan:2010rp}
T.~Padmanabhan, ``{Entropy density of spacetime and the Navier-Stokes fluid
  dynamics of null surfaces},'' {\em Phys.Rev.} {\bf D83} (2011) 044048,
  \href{http://www.arXiv.org/abs/1012.0119}{{\tt 1012.0119}}.

\bibitem{Kolekar:2011gw}
S.~Kolekar and T.~Padmanabhan, ``{Action principle for the Fluid-Gravity
  correspondence and emergent gravity},'' {\em Phys.Rev.} {\bf D85} (2012)
  024004,
\href{http://www.arXiv.org/abs/1109.5353}{{\tt 1109.5353}}.
%%CITATION = ARXIV:1109.5353;%%.

\bibitem{Damour:1979}
T.~Damour, ``{Quelques propri´et´es m´ecaniques, ´electromagn´etiques,
  thermodynamiques et quantiques des trous noirs (available at
  http://www.ihes.fr/∼damour/Articles/)},'' {\em Th`ese de doctorat
  d’´Etat, Universit´e Paris 6 .} (1979).

\bibitem{Damour:1982}
T.~Damour, ``{Surface effects in black hole physics},'' {\em Proceedings of the
  Second Marcel Grossmann Meeting on General Relativity} (1982).

\bibitem{Thorne}
K.~S. Thorne, R.~H. Price, and D.~A. MacDonald, ``{Black Holes: The Membrane
  Paradigm},'' {\em Yale University Press} (1986).

\bibitem{Kolekar:2011bb}
S.~Kolekar, D.~Kothawala, and T.~Padmanabhan, ``{Two Aspects of Black hole
  entropy in Lanczos-Lovelock models of gravity},'' {\em Phys.Rev.} {\bf D85}
  (2012) 064031,
\href{http://www.arXiv.org/abs/1111.0973}{{\tt 1111.0973}}.
%%CITATION = ARXIV:1111.0973;%%.

\bibitem{Gibbons:1976ue}
G.~Gibbons and S.~Hawking, ``{Action Integrals and Partition Functions in
  Quantum Gravity},'' {\em Phys.Rev.} {\bf D15} (1977)
2752--2756.
%%CITATION = PHRVA,D15,2752;%%.

\bibitem{gravitation}
T.Padmanabhan, {\em {Gravitation: Foundations and Frontiers}}.
\newblock Cambridge University Press, Cambridge, UK, 2010.

\bibitem{Ostrogradsky:1850}
M.~V. Ostrogradsky {\em Memoires de lAcademie Imperiale des Science de
  Saint-Petersbourg} {\bf 4} (1850) 385.

\bibitem{York:1972sj}
J.~York, James~W., ``{Role of conformal three geometry in the dynamics of
  gravitation},'' {\em Phys.Rev.Lett.} {\bf 28} (1972)
1082--1085.
%%CITATION = PRLTA,28,1082;%%.

\bibitem{Charap:1982kn}
J.~Charap and J.~Nelson, ``{Surface Integrals and the Gravitational Action},''
  {\em J.Phys.A:Math.Gen.} {\bf 16} (1983)
1661.
%%CITATION = PRINT-82-0726 (QUEEN MARY COLL) ETC.;%%.

\bibitem{Eddington:1924}
A.~Eddington, {\em {The Mathematical Theory of Relativity}}.
\newblock Cambridge University Press, Cambridge, UK, 2~ed., 1924.

\bibitem{Schrod:1950}
E.~Schrodinger, {\em {Space-time Structure, Cambridge Science Classics}}.
\newblock Cambridge University Press, Cambridge, UK, 1950.

\bibitem{Einstein:1955ez}
A.~Einstein and B.~Kaufman, ``{A new form of the general relativistic field
  equations},'' {\em Annals Math.} {\bf 62} (1955)
128--138.
%%CITATION = ANMAA,62,128;%%.

\bibitem{Einstein-Kauffman}
A.~Einstein and B.~Kaufman, ``{A New Form of the General Relativistic Field
  Equations},'' {\em Annals Math.} {\bf 62} (1955) 128--138.

\bibitem{Padmanabhan:2012bs}
T.~Padmanabhan, ``{Equipartition energy, Noether energy and boundary term in
  gravitational action},'' {\em Gen.Rel.Grav.} {\bf 44} (2012) 2681--2686,
\href{http://www.arXiv.org/abs/1205.5683}{{\tt 1205.5683}}.
%%CITATION = ARXIV:1205.5683;%%.

\bibitem{Majhi:2013jpk}
B.~R. Majhi and T.~Padmanabhan, ``{Thermality and Heat Content of horizons from
  infinitesimal coordinate transformations},''
\href{http://www.arXiv.org/abs/1302.1206}{{\tt 1302.1206}}.
%%CITATION = ARXIV:1302.1206;%%.

\bibitem{Padmanabhan:2013xyr}
T.~Padmanabhan and D.~Kothawala, ``{Lanczos-Lovelock models of gravity},''
\href{http://www.arXiv.org/abs/1302.2151}{{\tt 1302.2151}}.
%%CITATION = ARXIV:1302.2151;%%.

\bibitem{meaning}
A.~Einstein, {\em {The Meaning of Relativity}}.
\newblock Princeton University Press, 5~ed., 2004.

\bibitem{Kijowski:97}
J.~Kijowski, ``{A simple Derivation of Canonical Structure and quasi-local
  Hamiltonians in General Relativity},'' {\em Gen. Relat. Grav. Journal} {\bf
  29} (1997) 307--343.

\bibitem{Babak:1999dc}
S.~Babak and L.~Grishchuk, ``{The Energy momentum tensor for the gravitational
  field},'' {\em Phys.Rev.} {\bf D61} (1999) 024038,
\href{http://www.arXiv.org/abs/gr-qc/9907027}{{\tt gr-qc/9907027}}.
%%CITATION = GR-QC/9907027;%%.

\bibitem{MTW}
C.~W. Misner, K.~S. Thorne, and J.~A. Wheeler, {\em {Gravitation}}.
\newblock W. H. Freeman and Company, 3~ed., 1973.

\bibitem{Goldstein}
H.~Goldstein, C.~Poole, and J.~Safko, {\em {Classical Mechanics}}.
\newblock Pearson Education, 3~ed., 2007.

\bibitem{Palatini}
A.~Palatini, ``Deduzione invariantiva delle equazioni gravitazionali dal
  principio di hamilton,'' {\em Rend. Circ. Mat. Palermo} {\bf 43} (1919)
  203--212.

\bibitem{AE}
A.~Einstein, {\em Der Energiesatz in der allgemeinen Relativit{\"a}tstheorie},
  pp.~154--166.
\newblock Wiley-VCH Verlag GmbH \& Co. KGaA, 2006.

\bibitem{Dirac:1975}
P.~Dirac, {\em {General Theory of Relativity}}.
\newblock John Wiley and Sons, New York, 1975.

\bibitem{Medved:2004ih}
A.~Medved, D.~Martin, and M.~Visser, ``{Dirty black holes: Space-time geometry
  and near horizon symmetries},'' {\em Class.Quant.Grav.} {\bf 21} (2004)
  3111--3126,
\href{http://www.arXiv.org/abs/gr-qc/0402069}{{\tt gr-qc/0402069}}.
%%CITATION = GR-QC/0402069;%%.

\bibitem{Racz:1995nh}
I.~Racz and R.~M. Wald, ``{Global extensions of space-times describing
  asymptotic final states of black holes},'' {\em Class.Quant.Grav.} {\bf 13}
  (1996) 539--553,
\href{http://www.arXiv.org/abs/gr-qc/9507055}{{\tt gr-qc/9507055}}.
%%CITATION = GR-QC/9507055;%%.

\bibitem{Carter:1973}
B.~Carter, ``Republication of: Black hole equilibrium states part ii. general
  theory of stationary black hole states,'' {\em General Relativity and
  Gravitation} {\bf 42} (2010) 653--744.

\bibitem{Carroll:1997ar}
S.~M. Carroll, ``{Lecture notes on general relativity},''
\href{http://www.arXiv.org/abs/gr-qc/9712019}{{\tt gr-qc/9712019}}.
%%CITATION = GR-QC/9712019;%%.

\bibitem{Hollands:2006rj}
S.~Hollands, A.~Ishibashi, and R.~M. Wald, ``{A Higher dimensional stationary
  rotating black hole must be axisymmetric},'' {\em Commun.Math.Phys.} {\bf
  271} (2007) 699--722,
\href{http://www.arXiv.org/abs/gr-qc/0605106}{{\tt gr-qc/0605106}}.
%%CITATION = GR-QC/0605106;%%.

\bibitem{Morales}
E.~M. Morales, ``{On a Second Law of Black Hole Mechanics in a Higher
  Derivative Theory of Gravity},'' {\em
  http://www.theorie.physik.uni-goettingen.de/forschung/qft/theses/dipl/Morfa-Morales.pdf}
(2008).
%%CITATION = GR-QC/0605106;%%.

\bibitem{Arnowitt:1962hi}
R.~L. Arnowitt, S.~Deser, and C.~W. Misner, ``{The Dynamics of general
  relativity},'' \href{http://www.arXiv.org/abs/gr-qc/0405109}{{\tt
  gr-qc/0405109}}.

\bibitem{PhysRevD.2.2762}
B.~F. Schutz, ``Perfect fluids in general relativity: Velocity potentials and a
  variational principle,'' {\em Phys. Rev. D} {\bf 2} (Dec, 1970) 2762--2773.

\bibitem{Brown:1992kc}
J.~D. Brown, ``{Action functionals for relativistic perfect fluids},'' {\em
  Class.Quant.Grav.} {\bf 10} (1993) 1579--1606,
\href{http://www.arXiv.org/abs/gr-qc/9304026}{{\tt gr-qc/9304026}}.
%%CITATION = GR-QC/9304026;%%.

\bibitem{PhysRevD.86.087502}
O.~Minazzoli and T.~Harko, ``New derivation of the lagrangian of a perfect
  fluid with a barotropic equation of state,'' {\em Phys. Rev. D} {\bf 86}
  (Oct, 2012) 087502.

\end{thebibliography}\endgroup

\bibliographystyle{./utphys}

\end{document}